\documentclass[aps,prx,twocolumn,superscriptaddress,nofootinbib]{revtex4-1}
\usepackage[colorlinks=true,citecolor=blue,linkcolor=blue,urlcolor=blue]{hyperref}
\usepackage{amsmath, amssymb}
\usepackage{graphicx, nicefrac, textcomp}
\usepackage{mwe}
\usepackage{braket}
\usepackage{color}
\usepackage{dcolumn}
\usepackage[utf8]{inputenc}
\usepackage{float}
\usepackage{dsfont}
\usepackage{svg}
\usepackage[version=4]{mhchem}
\usepackage[toc,page]{appendix}
\usepackage{soul}

\begin{document}
\title{Neural-network-supported basis optimizer for the configuration interaction problem in quantum many-body clusters: Feasibility study and numerical proof}

\author{Pavlo Bilous}
\email{pavlo.bilous@mpl.mpg.de}
\affiliation{Max Planck Institute for the Science of Light, Staudtstraße 2, 91058 Erlangen, Germany}

\author{Louis Thirion}
\affiliation{Friedrich-Alexander-Universit\"at Erlangen/N\"urnberg, Department of Physics, Staudtstraße 7,  91058 Erlangen, Germany}

\author{Henri Menke}
\affiliation{Friedrich-Alexander-Universit\"at Erlangen/N\"urnberg, Department of Physics, Staudtstraße 7, 91058 Erlangen, Germany}

\author{Maurits W. Haverkort}
\affiliation{Heidelberg University, Institute for Theoretical Physics, Philosophenweg 19, 69120 Heidelberg, Germany}

\author{Adriana P\'alffy}
\affiliation{University of W\"urzburg, Institute of Theoretical Physics and Astrophysics,
 Am Hubland, 97074 W\"urzburg, Germany}

\author{Philipp Hansmann}
\email{philipp.hansmann@fau.de}
\affiliation{Department of Physics, Friedrich-Alexander-Universit\"at Erlangen/N\"urnberg, 91058 Erlangen, Germany}

\date{\today}

\begin{abstract}
A deep-learning approach to optimize the selection of Slater determinants in configuration interaction calculations for condensed-matter quantum many-body systems is developed.  We exemplify our algorithm on  the discrete version of the single-impurity Anderson model with up to 299 bath sites. Employing a neural network classifier and active learning, our algorithm enhances computational efficiency by iteratively identifying the most relevant Slater determinants for the ground-state wavefunction. We benchmark our results against established methods and investigate the efficiency of our approach as compared to other basis truncation schemes. Our algorithm  demonstrates a substantial improvement in the efficiency of determinant selection, yielding a more compact and computationally manageable basis without compromising accuracy. Given the straightforward application of our neural network-supported selection scheme to other model Hamiltonians of quantum many-body clusters, our algorithm can significantly advance selective configuration interaction calculations in the context of correlated condensed matter.
\end{abstract}

\maketitle

\section{Introduction}
\label{section:intro}

Strongly correlated quantum many-body systems pose great computational challenges throughout a plethora of research fields, including quantum chemistry, condensed matter, atomic, and nuclear physics. The difficulty arises from the exponential scaling of the Hilbert space with the total number of single-particle degrees of freedom, e.g., spin, orbital, lattice site, depending on the specific system. Numerical accurate solutions can be obtained by  configuration interaction (CI) methods, which express the fully interacting wave function of fermionic systems as a linear combination of Slater determinant basis states and compute the coefficients of this expansion by solving the Hamiltonian eigenvalue problem \cite{Szabo_ModnQChem_2012}. Unfortunately, CI becomes computationally impractical already for small molecules. For instance,  a full CI ground-state energy benchmark of the   N$_2$ molecule involved already approx.~$10^{10}$ Slater determinants \cite{ROSSI1999}. This figure demonstrates the "exponential wall" which makes full CI for larger molecules and solids impractical.

One possible strategy to tackle larger systems are so-called \emph{embedding techniques}. In a multi-tier scheme the full problem is subdivided into a smaller strongly correlated part, the quantum many-body cluster, which is coupled to a non-interacting bath. Examples in quantum chemistry are Multi-Configurational Self-Consistent Field methods (MCSCF) and a plethora of variants \cite{MCSCF01,MCSCF02,MCSCF03}. In condensed-matter physics,  the merger of density-functional theory with dynamical mean-field theory (DFT+DMFT) and its extensions \cite{dmft1,Georges1992,dmft2,cdmft1,cdmft2} belong to the most successful approximation schemes for predictions of strongly correlated materials.
For all embedding methods, the size of the quantum cluster might still be prohibitive for accurately capturing  essential quantum mechanical correlations. To this end, based on the observation that  only a small subset of the configurations contributes significantly to the description of the eigenfunctions \cite{Ivanic2001}, a large variety of selected CI methods have been developed, in particular in quantum chemistry, see for example Refs.~\cite{Huron1973,Greer1998} or \cite{Garniron2018,Tubman2020} and references therein. Selected CI approaches iteratively construct a compact wave function that captures the physics of the full CI wave function on the desired level of accuracy. 

At the same time, the past years have witnessed an impressive growth of machine learning (ML) applications to quantum chemistry and computational materials science. Starting from quantum chemistry, ML has been successfully applied to explore the configuration space in CI and to construct the wave function keeping just the most important Slater determinants \cite{Coe_MLCI_JChemTC_2018,Jeong_ALCI_JChemTC_2021, Chembot, RLCI, Herzog2023}.  In the original approach by Coe \cite{Coe_MLCI_JChemTC_2018,Coe_JChemTC_2019}, a regression neural network (NN) was explicitly computing the coefficient of each atomic configuration, allowing to iteratively construct the wave function. Later approaches perform classification directly, as the NN decides whether configurations are important or unimportant, without actually predicting their coefficients \cite{Jeong_ALCI_JChemTC_2021}. Related approaches have been developed recently also for the structure and dynamics of light nuclei \cite{Molchanov2022} and for accurate atomic structure calculations \cite{MLGRASP}.

In this work, we develop a related strategy using the active learning approach to iteratively select the most relevant Slater determinants and construct an approximate wave function for a class of effective second quantization model Hamiltonians.
Our aim is to boost the performance of selected CI methods for quantum many-body systems and quantum clusters in the realm of solid-state physics. We apply our method to one of the prototypical models for strongly correlated electron systems, the single-impurity Anderson model (SIAM) \cite{And1961} and demonstrate its feasibility by performing ground-state calculations. While certain computation parameters presented here are specific for SIAM, our algorithm is designed in a more general manner and in principle applicable to any quantum many-body Hamiltonian of a finite size cluster. 

More specifically, we apply our NN algorithm to the SIAM model in the star geometry \cite{Aichhorn2011} for several bath site numbers up to $N_{\rm bath}=299$.  The NN receives the input information on Slater Determinants (SDets) encoded as binary populations of impurity and bath sites separately for spin-up and spin-down states. Training is performed in several stages on smaller random selections, using the results of exact diagonalization provided by the software package \textsc{Quanty} \cite{Lu2014}. The ground state wave function is constructed iteratively and convergence is monitored by the variance of the ground state energy. By evaluating several physical observables such as electron density, double occupancy, and the static magnetic susceptibility at zero temperature, we successfully benchmark our results against available SIAM data presented in Ref.~\cite{Aichhorn2011}. The performance of our NN-supported selective CI algorithm is analysed quantitatively. We compare the accuracy of our calculations as a function of Hilbert space to results obtained by other non NN-based truncation schemes \cite{Lu2014, Cao2021}. Our results demonstrate the superior efficiency (accuracy vs. determinant basis size) of the NN-selected basis and its potential for both increasing the best accuracy or allowing for larger bath site numbers. 

We note that, as in quantum chemistry, ML has been an active player in tackling the many-body problem in condensed-matter systems in various approaches \cite{Carleo2019,  Dey2023}. Carleo and Troyer showed how a many-body wave function can be approximated by encoding it into a NN \cite{Carleo2017}. This approach was used to calculate the ground state in the one-dimensional transverse-field Ising model, the antiferromagnetic Heisenberg model \cite{Carleo2017}, and other effective models of solid state physics \cite{SchmittRamsDziarmagaetal.2022,  Saito2017} and quantum chemistry \cite{Choo2020}. With DFT alone ML has been employed in different approaches, e.g., to optimize effective density functionals \cite{Snyder2012,Seino2018,Dick2020}, to construct approximations to the local density map \cite{Brockherde2017,Moreno2020}, and to construct optimized pseudo-potentials in plane-wave based DFT codes \cite{DFTpseudoml1,DFTpseudoml2,DFTpseudoml3}. In the context of the matrix product state ansatz,  ML-based optimization has been tested for the Heisenberg Hamiltonian for one-dimensional lattices with nearest neighbor interactions \cite{Ghosh2023}. ML was also used in the context of DMFT for the SIAM model \cite{Arsenault2014,Rigo2020,Sheridan2021,Sturm2021,Walker2022}. Our newly conceptualized selective CI approach for quantum clusters presents a complementary strategy in this vibrant field and might, in combination with the other methods, help to improve computationally affordable calculations for correlated electron systems beyond the single-particle approximation.

The paper is structured as follows. Section \ref{section:algo} presents our algorithm, the computational protocol, and the description of the NN architecture concluding with remarks on possible optimization. The application to the SIAM model follows in Section \ref{section:application}. Here we present numerical results, benchmarks, and discuss the performance of our NN-assisted selective CI. The paper closes with summary and conclusions in Section \ref{section:summary}. 

\begin{figure*}[ht]
     \begin{center}
     \includegraphics[width=0.9\textwidth]{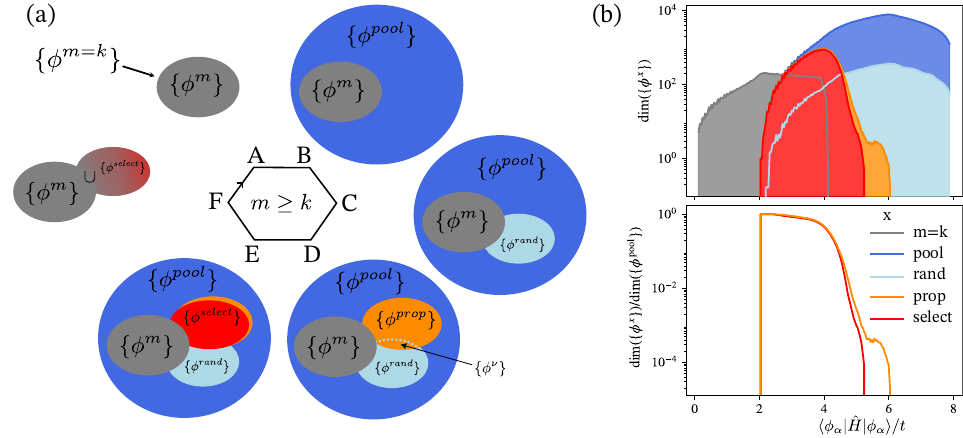}
     \end{center}
     \caption{
     (a): Sketch of the six stages \textit{A} to \textit{F} in our iterative algorithm after $k$ initialization steps. We indicate the definition of the different sets (set at iteration $m$, pool, random sample, proposed, and selected) of Slater determinants (see text for definitions and further explanation). (b): Distribution functions of the corresponding sets $\{ \phi^\text{x} \}$  over their  energy matrix elements in the first ML iteration of our application to the SIAM model discussed in Section \ref{section:application} for $N_\text{bath}=89$, $U=4$\,eV and $\varepsilon_\text{imp}=0$\,eV. Note that overlapping energy distributions do not imply overlapping sets.
     }
     \label{fig:alg}
\end{figure*}

\section{Algorithm}
\label{section:algo}
ML approaches are typically distinguished into supervised learning,  unsupervised learning and reinforcement learning paradigms \cite{pml1Book}. Starting from the deep-learning algorithm developed for atomic structure calculations in Ref.~\cite{MLGRASP}, we use the so-called \emph{active learning} (AL) approach, which does not strictly fall into one of these three standard categories. Specifically, we design a NN classifier that is trained in a supervised way,  whereas the data are not given a priori (as in the standard supervised learning),  but are produced \textit{actively} during interaction with the ``environment''.  Decisions in this interaction process are in turn based on the results of the previous training iterations. These are common features of AL and reinforcement learning.  However, in AL there is no explicit reward to be maximized. Instead, when applied to solution of the CI problem, the observable of interest (in the following this will be the ground state energy, but it could be any other observable) is monitored as a convergence criterion. In the following we define the relevant quantities for our procedure, describe the computational protocol, provide details of the NN architecture, and finally list potential adjustments to adapt our approach to different cluster models.\\

\paragraph{Definitions.}
The exact ground state of a quantum many body Hamiltonian  
\begin{equation}
\hat{H}  \ket{\Psi_\text{gs}} = E_\text{gs}  \ket{\Psi_\text{gs}}
\end{equation}
can be expanded in SDets $\phi$ which provide an orthonormal basis of the full Hilbert space ${\cal H}$ with $\braket{ \phi_i| \phi_j}=\delta_{i,j}$ and ${\cal H}^\text{full}=\text{span}\left(\{ \phi \}\right)$. The expansion reads
\begin{equation}
\ket{\Psi_\text{gs}} = \sum^{N}_\alpha c_\alpha \ket{\phi_\alpha}\, ,
\end{equation}
with $N\equiv\text{dim}\left({\cal H}^\text{full}\right)$. 
We aim to approximate the exact ground state $\ket{\Psi_\text{gs}}$ by searching for the most relevant subspace of ${\cal H}^s \subset {\cal H}^\text{full}$ with $\text{dim}\left({\cal H}^s\right)=N^s\ll N$  for the calculation of our ``target quantity'', the ground state energy $E_\text{gs}$. The approximate wave function can be written as
\begin{equation}
\ket{\Psi_\text{gs}} \approx \sum^{N^s\ll N}_\alpha c_\alpha \ket{\phi_\alpha}.
\end{equation}
For the basis state selection procedure we employ a NN as described in the following. \\

\paragraph{Computation protocol.}
For a given Hamilton operator $\hat{H}$ our computation starts with an initial basis $\{ \phi^\text{init} \}$ spanning a Hilbert space ${\cal H}^\text{init}=\text{span}\left(\{ \phi^\text{init} \}\right)$ with a dimension which might be as small as one. 

Next, we define an \emph{extension operator} $\hat{\cal O}$ and a fixed number of initialization extension steps $k$. Acting with $\hat{\cal O}$ on the elements of $\{ \phi^\text{init} \}$ projects out of ${\cal H}^\text{init}$ and onto new basis SDets. After $m\leq k$ applications we write 
\begin{equation}
\{ \phi^m \} = \hat{\cal O}^m \cdot \{ \phi^\text{init} \} \; .
\end{equation}
On a given basis $\{ \phi^m \}$, the Hamilton operator can be written in form of a matrix
\begin{equation}
H^m_{ij}=\braket{ \phi^m_i|\hat{H}| \phi^m_j}
\end{equation}
for which we find the lowest eigenvalue \mbox{$E^m_\text{gs}=\braket{ \Psi^m_\text{gs}|\hat{H}|\Psi^m_\text{gs}}$} with the corresponding ground state
\begin{equation}
\ket{\Psi^m_\text{gs}} = \sum^{\text{dim}\left({\cal H}^m\right)}_\alpha c_\alpha \ket{\phi^m_\alpha}.
\end{equation}
In order to quantify the accuracy of $\ket{\Psi^m_\text{gs}}$ as an approximation to the exact ground state we evaluate the variance
\begin{equation}
\label{eq:var}
    (\sigma^m_\text{gs})^2 \equiv \braket{ \Psi^m_\text{gs}|\left(\hat{H}-E^m_\text{gs}\right)^2| \Psi^m_\text{gs}}
\end{equation}
as a control parameter that vanishes upon convergence to the exact ground state. Such converge is, however, only possible if $\hat{\cal O}^m \cdot \{ \phi^\text{init}\}$ has a \emph{finite projection on the exact solution}. Neither $\{ \phi^\text{init}\}$ nor $\hat{\cal O}$ should, hence, be chosen too restrictively. (A reasonable choice could be to take the mean-field solution of $\hat{H}$ as $\{ \phi^\text{init}\}$ and $\hat{\cal O}=\hat{H}$.) As $\text{dim}\left({\cal H}^m\right)$ grows rapidly, the bases $\{ \phi^m \}$ quickly become too large to handle. 

Therefore, we interrupt the extension procedure after $k$ steps and begin with the NN-supported selection process which we sketch visually in Fig.~\ref{fig:alg}. The basis which is created in the initial process becomes the so-called primary set: $\{ \phi^\text{prim} \}\equiv \{ \phi^{m=k} \}$, see (A) in Fig.~\ref{fig:alg}. The primary set is always included in the wave function and never exposed to the NN. It is fixed and remains unchanged for the rest of the computation.

Next, for all iterations with $m\geq k$ we act with the extension operator on the basis set $\{\phi^m\}$ to obtain the \emph{pool} - see (B) in Fig.~\ref{fig:alg}:
\begin{equation}
\{ \phi^\text{pool}\} \equiv \hat{\cal O} \cdot \{ \phi^m\}\, .
\end{equation}
Instead of proceeding with the entire pool we define a fraction $\nu < 1$ of basis SDets which should be selected by the NN. To train our NN, we take a randomly sampled set of size $R$ from the pool
\begin{equation}
\{ \phi^\text{rand}\} \subset \left(\{ \phi^\text{pool}\}\setminus \{ \phi^\mathrm{m} \}\right)
\end{equation}
and compute the ground state $\Psi^\text{rand}_\text{gs}$ on the union $\{ \phi^m \} \cup \{ \phi^\mathrm{rand} \}$ (see stage C in Fig.~\ref{fig:alg}).
The importance of the basis determinants in the randomly sampled set is quantitatively determined by their coefficients in the calculated ground state
\begin{equation}
c^\text{rand}_\alpha\equiv\braket{ \Psi^\text{rand}_\text{gs}|\phi^\text{rand}_\alpha}\, .
\end{equation}
The SDets in the sampled set whose absolute values of the coefficient $|c^\text{rand}_\alpha|$ are larger than a positive valued cutoff parameter $c^\text{rand}_\text{min}$ are labelled as ``important'' and form the set $\{ \phi^\nu\} $; the remaining ones are ``unimportant''. Here, $c^\text{rand}_\text{min}$ is chosen such that the dimension of the set $\{ \phi^\nu\} $ is approximately the fraction $\nu$ of the set $\{ \phi^\text{rand}\}$. We assume that the same $c^\text{rand}_\text{min}$ splits also the whole pool $\{ \phi^\text{pool}\}$ into ``important'' and ``unimportant'' parts in approximately the same proportion as $\{ \phi^\text{rand}\}$. This condition is well satisfied if the chosen primary set is large enough to guarantee that the basis expansion coefficients $c_\alpha$ for the same SDets do not depend strongly on presence of other SDets from the pool in the wave function. Also, this condition does not pose considerable restrictions on our computations, since we are interested not directly in the values of $c_\alpha$ but their value with respect to the chosen cutoff $c^\text{rand}_\text{min}$.

The NN training set for an arbitrary NN-supported iteration $m \ge k$ is formed as
\begin{equation}
\{ \phi^\text{train}\} = \{ \phi^\text{rand}\} \cup \left(\{ \phi^m\}\setminus \{ \phi^\text{prim} \}\right)\,.
\end{equation}
Note that in the very first NN-supported iteration, i.e. for $m = k$, we have $\{ \phi^\text{train}\} = \{ \phi^\text{rand}\}$. The trained NN proceeds to classify in the next step each SDet in $\{ \phi^\text{pool}\}\setminus \{ \phi^\mathrm{train} \}$ as ``important'' or ``unimportant''. At the end of this procedure, the NN proposes a set of important SDets which unified with $\{ \phi^\nu\} $ form the set $\{ \phi^\mathrm{prop}\}$. 
If $c^\text{rand}_\text{min}$ was chosen appropriately, the number of proposed important SDets should be approximately the fraction $\nu$ of the number of determinants in the pool (stage D in Fig.~\ref{fig:alg}). We stress that the parameters $\nu$ and $R$ involved in this step are chosen by the user.

In the second to last step we then diagonalize the Hamiltonian once more on the union 
$\{ \phi^m \}\cup \{ \phi^\mathrm{prop}\}$. Finally, we select those proposed 
SDets for which we find the coefficients
\begin{equation}
c^\text{prop}_\alpha\equiv\braket{ \Psi^\text{prop}_\text{gs}|\phi^\text{prop}_\alpha}
\end{equation}
with larger absolute values than the cutoff $c^\text{rand}_\text{min}$, so that (stage E in Fig.~\ref{fig:alg})

\begin{equation}
\{\phi^\text{select}\} = \{ \phi^\text{prop}_\alpha \in \{\phi^\text{prop}\}\mid c^\text{rand}_\text{min} \leq |c^\text{prop}_\alpha| \}    \, .
\end{equation}
The resulting basis of SDets at the end of the iteration is then
\begin{equation}
\{ \phi^{m+1} \} = \{ \phi^m \} \cup \{ \phi^\mathrm{select} \}\,.
\end{equation}

\begin{figure}[t]
     \begin{center}
     \includegraphics[width=\columnwidth]{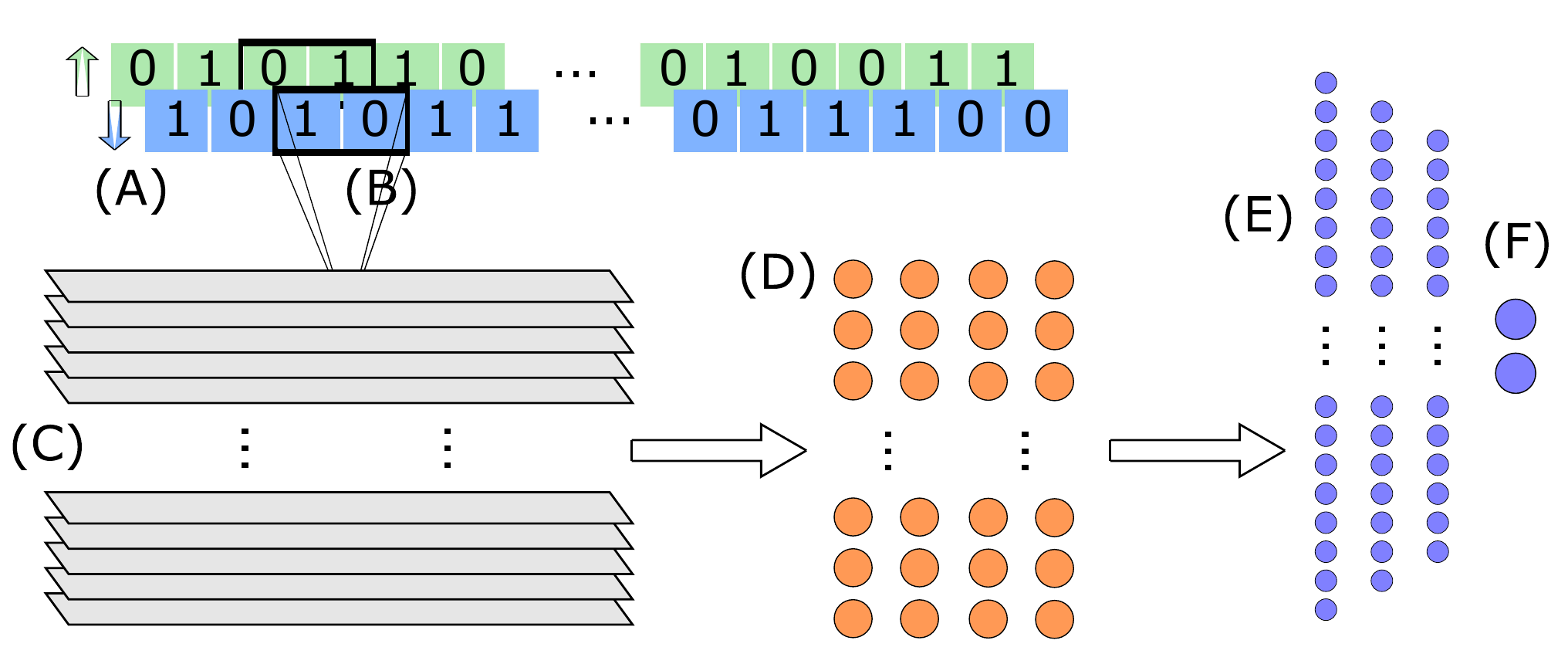}
     \end{center}
     \caption{Architecture of the convolutional NN used in this work. A candidate SDet is given as input (A) and eventually classified as important or unimportant (F). See text for further explanations. }
     \label{fig:nn_architecture}
\end{figure}

\paragraph{NN architecture.} 
In Fig.~\ref{fig:nn_architecture}, the NN architecture implemented in this study is illustrated. The NN processes the input, which is a candidate SDet $\ket{\phi_\alpha}$, represented in a (spin-orbital) occupation number format as a string of 0s and 1s (A). This input is split into two channels for spin-up and spin-down orbitals. The data then pass through a convolutional filter kernel of size 2 (B), generating 64 feature maps (C). These maps are subsequently processed by a purely local kernel, resulting in 4 output channels (D). These channels are flattened and forwarded to a dense block (E) which ends with an output layer consisting of two neurons (F). These neurons classify the input SDet as ``important" or ``unimportant" using a softmax activation function, which ensures the outputs are normalized to lie between 0 and 1 and sum up to 1.

Throughout the network, the rectified linear unit (ReLU) is employed as the activation function for hidden layers, and the network’s performance is evaluated using categorical cross-entropy. The Adam algorithm is used for training, which terminates after no improvement is observed over three consecutive epochs, a method known as ``early stopping with patience". This architecture, which is of the convolutional type \cite{Goodfellow2016}, has been demonstrated to be effective for solving the configuration interaction (CI) problem in atomic structure computations \cite{MLGRASP}. The implementation was carried out using the \textsc{Python} library \textsc{TensorFlow}  \cite{TensorFlow2015}.\\

\paragraph{Remarks and possibilities for optimization.}
 We note that in the current procedure the NN is retained for the subsequent cycles keeping the memory of previous training instances. Next, it is important to stress that discarded determinants are never ``lost''. In later cycles, after additional extension steps, they might be re-selected. Furthermore, a quantitative metric for the convergence has to be defined for the specific problem/model that is treated. For our SIAM feasibility study we chose the ground state energy (and its variance). 
 
 \begin{figure}[t]
     \begin{center}
     \includegraphics[width=\columnwidth]{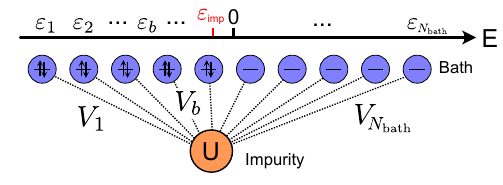}
     \end{center}
     \caption{Sketch of the SIAM model in star geometry. The impurity is connected to $N_\mathrm{bath}$ non-interacting bath sites with energies $\varepsilon_b$ and  hybridization amplitudes $V_b$, with $b=1,\ldots ,N_\mathrm{bath}$. The impurity is characterized by energy $\varepsilon_{\rm imp}$ and the particle-hole symmetric onsite interaction $U$.}
     \label{fig:SIAM_sketch}
\end{figure}

 There are a number of possibilities to modify our algorithm to optimize the procedure for other cluster models including finite size chunks of Hubbard or Heisenberg lattices, full multiplet ligand field theory clusters \cite{Haverkort_2012}, or even ab initio derived (i.e. with DFT or Hartree-Fock) Hamiltonians for modeling molecules, for instance: 
\begin{itemize}
    \item NN architecture: The user can modify the convolutional block [the kernel size (B), the number of the feature maps (C)], the dense block (E) and the coupling between them (D). In principle, the convolutional NN can be replaced by any other architecture complying with the input and output structure. Also other classifiers can be explored, e.g. support vector machines as in Ref. \cite{Chembot}.
    \item Training parameters: We trained our model using the Adam algorithm which is often the default choice for training of neural networks. Depending on the specific computation, other optimizers can be employed. The approach of early stopping with patience used here can be also adjusted or replaced. For more information on neural networks see. e.g. Ref. \cite{Goodfellow2016}. See also e.g. Ref. \cite{HandsOnML} for other ML approaches.
    \item Extension steps: In our current scheme we expand using the full Hamiltonian as extension operator. In larger models with more complex interactions this could be changed and modified extension steps (truncated interactions, etc.)  are possible.
    \item Size of the training set: The size of the training set (in our case determined by a heuristic function dependent on the pool size) needs to be chosen carefully as it should be small enough to not create a bottleneck for the procedure but remain representative of the pool. In order to systematically determine the optimal fraction, a careful analysis of the NN performance (fraction of false positive- and false negative selections) can be a valid strategy. 
    \item Size of selected fraction: The number of selected determinants changes the slope of the energy convergence and can be tuned accordingly, i.e. the fraction size can be decreased in order to maximize efficiency.  
\end{itemize}

Depending on the specific target model, other strategies to optimize convergence might be successful, for 
instance, pre-convergence in restricted subspaces of the full Hilbert space.

\begin{figure}[t]
    \centering
    \includegraphics[width=\columnwidth]{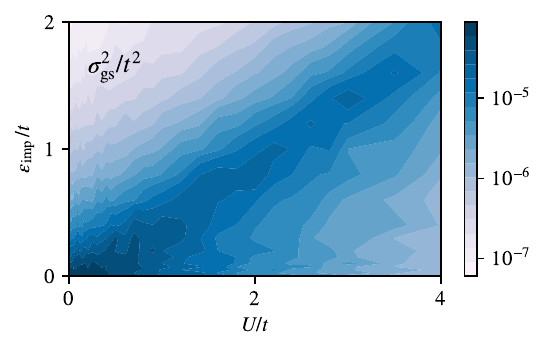} 
    \caption{Energy variance of the ground state energy in the $(\varepsilon_\text{imp}, U)$-plane on logarithmic color scale.}
    \label{fig:varE}
\end{figure}

\section{Application to SIAM model}
\label{section:application}
We now apply the NN algorithm to find the ground state of SIAM in the star geometry \cite{Aichhorn2011} which we sketch in Fig.~\ref{fig:SIAM_sketch}. The SIAM Hamilton operator reads
\begin{equation}
\label{eq:Hamiltonian}
\begin{split}
    \hat{H}_\text{SIAM}&=\varepsilon_\text{imp} \;\hat{n}^\text{imp}_\text{tot} + U\;\left(\hat{n}^\text{imp}_{\uparrow}-\frac{1}{2}\right)\left(\hat{n}^\text{imp}_{\downarrow}-\frac{1}{2}\right)\\ &+ \sum_{\sigma\in\{\uparrow,\downarrow\}}\sum_{b=1}^{N_\text{bath}} \left(\varepsilon_b \hat{n}^b_\sigma +  V_{b}(c_{\text{imp},\sigma}^\dagger c_{b,\sigma} + h.c.)\right)\, ,
\end{split}
\end{equation}
where $c^\dagger_{\alpha,\sigma}$ and $c_{\alpha,\sigma}$ are fermionic creation and annihilation operators, $\hat{n}^\text{imp}_\sigma\equiv c^\dagger_{\text{imp},\sigma} c_{\text{imp},\sigma}$  ($\hat{n}^b_\sigma\equiv c^\dagger_{b,\sigma} c_{b,\sigma}$) are the occupation operators of the impurity (bath) sites and $\hat{n}^\text{imp}_\text{tot}=\hat{n}_\uparrow^\text{imp}+\hat{n}_\downarrow^\text{imp}$ the total impurity occupation. The parameters of the model are the number of non-interacting bath sites $N_\text{bath}$, the onsite energies of the impurity $\varepsilon_\text{imp}$ and the bath sites $\varepsilon_\text{b}$, the hybridization amplitudes $V_b$, and $U$ as the particle-hole symmetric onsite interaction on the impurity site. Indeed the parameters can be chosen in a way that the star geometry maps directly to a 1D chain i.e., an impurity site coupled to the first site of a 1D bath chain with hybridization $V$ with constant nearest neighbor hopping $t$ \cite{Aichhorn2011}. Accordingly, we set 
\begin{equation}
\begin{split}
    \varepsilon_b &= -2 t \cos{\left(\frac{b\pi}{N_\mathrm{bath}+1}\right)}\, ,
    \\
    V_b &= V\sqrt{\frac{2}{N_\mathrm{bath}+1}}\sqrt{1-\left(\frac{\varepsilon_b}{2t}\right)^2}\, ,
\end{split}
\end{equation}
with bath site index $b$ running from 1 to $N_\text{bath}$, and fixed values of $V=0.1$ eV and $t=1.0$ eV as in \cite{Aichhorn2011}. In the following, we give all energies in units of $t=1.0$ eV. Moreover, we restrict our  calculations to an odd number of bath sites (such that there is always a bath site at $\varepsilon_{b}=0$) and half-filling, i.e. $N_{e}=N_\text{bath}+1$.

\begin{figure}[t]
    \centering
    \includegraphics[width=\columnwidth]{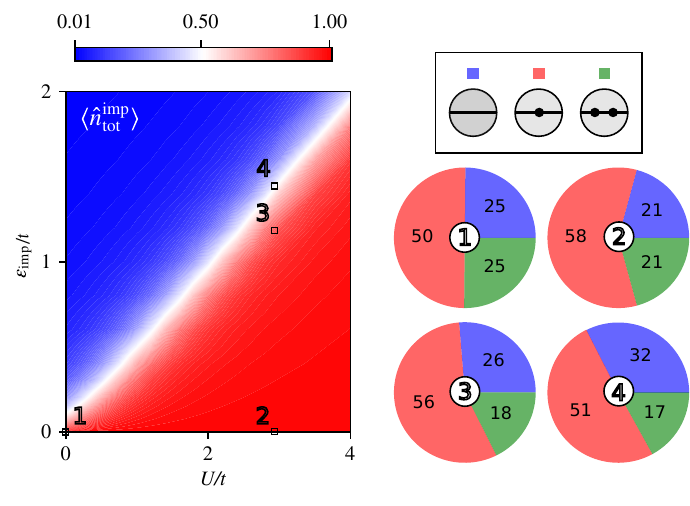} 
    \caption{Left panel: $\varepsilon_\text{imp}$ vs. $U$ phase diagram for the impurity density  $\langle n^\text{imp}_\text{tot}\rangle$. Right panel: Percentage contributions of empty, singly-, and doubly occupied configurations in the ground state at four selected points marked in the phase diagram.}
    \label{fig:density}
\end{figure}

For the calculations of physical observables presented in the next section we have fixed the number of bath sites in Eq.~\eqref{eq:Hamiltonian} to $N_\text{bath}=89$. In this way we were able to perform a large number of CI calculations to span a phase diagram for various values of impurity energy $\varepsilon_\text{imp}$ and interaction $U$. All calculations were started from the same spin degenerate zero-hybridization ground state  

\begin{equation} 
\label{eq:phi_init}
\{\phi_{S=0}^{\text{init}}\}= \{ c^{\dagger}_{\text{imp},\sigma} c^{\dagger}_{\text{b},\bar{\sigma}}\prod_{b}^{\varepsilon_{b}<0}c^{\dagger}_{b,\uparrow}c^{\dagger}_{b,\downarrow}|0\rangle \}
\end{equation}
(with total spin quantum number $S=0$ and $\bar{\sigma}\equiv-\sigma$). The procedure was then restricted to \emph{two initialization (extension) steps} followed by \emph{eight NN-supported iterations}. Due to the fixed number of steps, the accuracy [given by \eqref{eq:var}] of the obtained ground state varies in the phase diagram as can be seen in Fig.~\ref{fig:varE}. While there is a difference of over three orders of magnitude in the reached variance, the plot shows clearly that we are below the threshold of $\sigma_\mathrm{gs}^2<10^{-4}$ for all parameters. We note in passing, that the specific $\varepsilon_\text{imp}$, $U$ dependence of $\sigma_\mathrm{gs}^2$ is mostly caused by the ``distance" of our starting basis to the exact result, i.e., the number of neccessary extension steps. This is most visible for $\varepsilon_\text{imp}=U=0$\,eV where $\{\phi_{S=0}^{\text{init}}\}$ given by \eqref{eq:phi_init} presents a very poor approximation to the actual ground state.

\subsection{Results I: Observables}
\label{section:results1}

In this section we present numerical results for physical observables including the impurity density, the double occupancy, and the static magnetic susceptibility. The zero temperature observables ${\cal O}$ were computed by evaluating the expectation value 
\begin{equation}\label{eq:exp_val}
{\cal O}=\braket{ \Psi_\text{gs}| \hat{\cal O} | \Psi_\text{gs}} 
\end{equation}
with the ground state wavefunction $\Psi_\text{gs}$ obtained with our NN CI algorithm. The results were then compared to benchmark data for the SIAM problem presented in Nuss \emph{et al.}  \cite{Aichhorn2011}. 

\begin{figure}[t]
    \centering
    \includegraphics[width=\columnwidth]{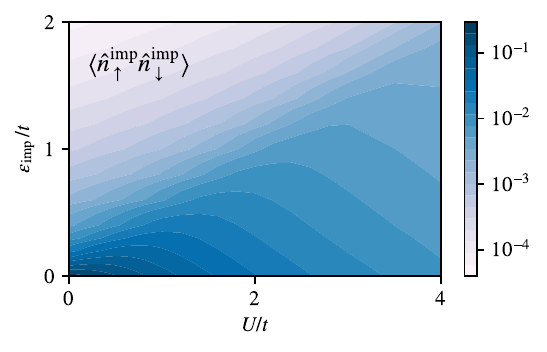} 
    \caption{$\varepsilon_\text{imp}$ vs. $U$ phase diagram for the impurity  double occupancy plotted on a logarithmic color scale.}
    \label{fig:double_occ}
\end{figure}

We start with the total impurity density $\langle \hat{n}^\text{imp}_\text{tot}\rangle$ as a function of the interaction $U$ and $\varepsilon_\text{imp}$ which we plot in the left panel of Fig.~\ref{fig:density}. Particle-hole symmetry of bath and the hybridization function as well as impurity interaction leads to the relation
\begin{equation}
\label{eq:p-h-sym} 
\langle \hat{n}^{\text{imp}}_\sigma\rangle(\varepsilon_\text{imp})= 1- \langle \hat{n}^{\text{imp}}_\sigma\rangle(-\varepsilon_\text{imp})
\end{equation}
so that all information is contained in the plot for $\varepsilon_\text{imp} >0$\,eV.
The data agrees well with Fig.~12 in Ref.~\cite{Aichhorn2011}. At $\varepsilon_\text{imp}=0$\,eV the impurity is exactly half filled for all values of $U$, while in the limit of large $\varepsilon_\text{imp}$ it is empty. As we have access to the full many-body wave function of the ground state we can resolve the result further in terms of contributing configurations. The pie charts in the right panels of Fig.~\ref{fig:density} show the composition of the ground state at four selected points in the phase diagram. At $U=0$\,eV and impurity half-filling (point 1), we find the expected equal distributions over the four possible impurity configurations (up, down, empty, full). In the large $U$ limit (point 2) one can clearly see how the distribution "sharpens" to the singly occupied impurity determinants for larger values of $U$ as double occupations of electrons and holes become energetically increasingly expensive. This trend is also true for finite $\varepsilon_\mathrm{imp}$ which, however, breaks the particle-hole symmetry of the distribution. 
Indeed the configuration composition visualized in the pie charts is also directly reflected by the two-particle observable measuring the \emph{double occupancy} $\langle\hat{n}^\text{imp}_\uparrow\hat{n}^{\text{imp}}_\downarrow\rangle$. As for the total impurity density, we can exploit the following symmetry relation with respect to $\varepsilon_\text{imp}$:
\begin{align}
\langle\hat{n}^{\text{imp}}_\uparrow\hat{n}^{\text{imp}}_\downarrow\rangle (-\varepsilon_\text{imp}) = &\langle\hat{n}^{\text{imp}}_\uparrow\hat{n}^{\text{imp}}_\downarrow\rangle (\varepsilon_\text{imp}) \nonumber\\
+ 1 - &\langle\hat{n}^{\text{imp}}_\uparrow + \hat{n}^{\text{imp}}_\downarrow\rangle (\varepsilon_\text{imp})
\end{align}
and show the plot of the double occupancy only for positive $\varepsilon_\text{imp}$ in Fig.~\ref{fig:double_occ}.

\begin{figure}[t]
     \begin{center}
     \includegraphics[width=0.9\columnwidth]{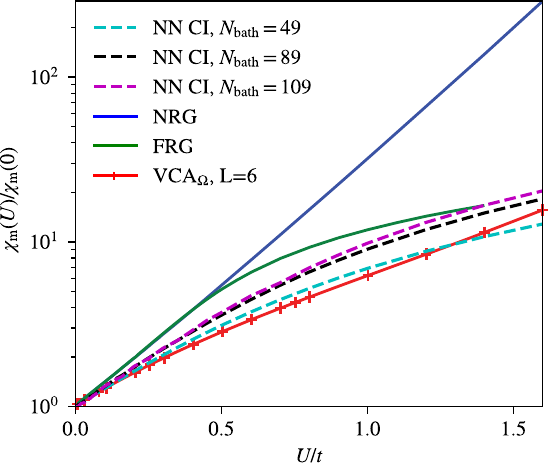}
     \end{center}
     \caption{Static magnetic susceptibility $\chi_\mathrm{m}$ as a function of impurity interaction $U$ at $\varepsilon_\text{imp}=0$\,eV. The NN CI results (dashed lines) are compared to VCA \cite{Aichhorn2011}, FRG\cite{Karrasch2008}, and (numerically exact) NRG\cite{Karrasch2008} benchmarks.
     }
     \label{fig:Susceptibility_Aichhorn}
\end{figure}

The suppression of empty/doubly occupied configurations upon increasing $U$ corresponds to a reduction of the double occupancy as discussed above. The limits for very large $\varepsilon_\text{imp}$ are trivial since as $\langle\hat{n}^\text{imp}_\text{tot}\rangle\rightarrow 0$ also  $\langle\hat{n}^\text{imp}_\uparrow\hat{n}^{\text{imp}}_\downarrow\rangle \rightarrow 0$. While being useful for the analysis of the many-body wave function, the double occupancy is not directly accessible in an experiment. Therefore, we also consider the (closely related) static magnetic susceptibility $\chi_\mathrm{m}$. In order to compute $\chi_\mathrm{m}$ we consider external fields which are much smaller than any other energy scale in the system, such that we are in a linear response regime and the relation 
\begin{equation}
    \chi_\mathrm{m} \approx \frac{ \langle\hat{S}_z\rangle}{B_z}
\end{equation}
holds. Here the expectation value of the $z$-component of the total spin operator $\langle\hat{S}_z\rangle$ on the right hand side is converged with our procedure in the presence of an external magnetic field $B_z$.

\begin{figure}[t]
    \centering
    \includegraphics[width=\columnwidth]{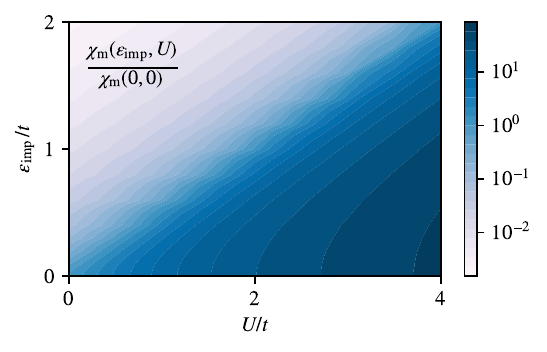} 
    \caption{$\varepsilon_\text{imp}$ vs. $U$ phase diagram for the static magnetic susceptibility $\chi_\text{m}$ for the SIAM with $N_\text{bath}=89$ on a logarithmic color scale.}
    \label{fig:Susceptibility}
  \end{figure}
  
In Fig.~\ref{fig:Susceptibility_Aichhorn} we compare  the NN CI susceptibility  obtained for $N_\text{bath}\in \{49,89,109\}$ (at $\varepsilon_\text{imp}=0$) with the variational cluster approximation (VCA$_\Omega$) for a system size $L=6$ (see \cite{Aichhorn2011} for details), functional renormalization group (FRG) \cite{Karrasch2008}, and the numerically exact zero temperature result of numerical renormalization group (NRG) \cite{Karrasch2008}. Clear discrepancies can be observed between the numerically exact NRG benchmark (blue line) and the NN CI results.  While at small $U$ we indeed observe the expected exponential increase of $\chi_\mathrm{m}(U)$, at larger $U$ the NN CI data deviates quantitatively and qualitatively (loss of exponential behaviour) from the exact result. Moreover, this seems to be at odds with the smaller variance $\sigma_\mathrm{gs}^2$ at large $U$ (see Fig.~\ref{fig:varE}) which suggests a \emph{better} accuracy of our ground state in this parameter regime. The seeming contradiction is, however, quickly resolved when we compare the NN CI data for different bath-site numbers $N_\mathrm{bath}$ (filled circles in Fig.~\ref{fig:Susceptibility_Aichhorn}). The trend towards the exact result with increasing $N_\mathrm{bath}$ indicates clearly that the problem lies in the bath discretization and not in the convergence of the CI procedure. At the considered cluster sizes we reproduce the exact exponential behaviour at smaller interaction values and systematically improve with the control parameter $N_\mathrm{bath}$. Furthermore, our results perform well in comparison to VCA with system size $L=6$ \cite{Aichhorn2011} 
and the weak-coupling FRG results \cite{Karrasch2008}.  

In Fig.~\ref{fig:Susceptibility} we plot $\chi_\mathrm{m}$ also for nonzero values of $\varepsilon_\text{imp}$ (for $N_\text{bath}=89$). Due to the particle-hole symmetry of $\hat{S_z}$ we show only the region $\varepsilon_\text{imp}\geq 0$. The increase of $\chi_\mathrm{m}$ at $\varepsilon_\text{imp}=0$ for increased values of $U$ reflects the correlation driven formation of larger local magnetic moments. This could have been anticipated from the drop of the impurity double occupancy on the $\varepsilon_\text{imp}=0$ line (see Fig.~\ref{fig:double_occ}). However, in the low/high impurity density limits, i.e. $\varepsilon_\text{imp}\neq 0$ , the effect of $U$ becomes negligible so that $\chi_\mathrm{m}(\varepsilon_\text{imp}\gg0, U) \approx \chi_\mathrm{m}(\varepsilon_\text{imp} \gg 0, U=0)$.
Closing the discussion on the magnetic susceptibility we emphazise once more, that the phase diagram of $\chi_\mathrm{m}(\varepsilon_\mathrm{imp},U)$ in Fig.~\ref{fig:Susceptibility} each NN CI curve in Fig.~\ref{fig:Susceptibility_Aichhorn} should  be understood as the converged result for the discrete SIAM Hamiltonian for the given $N_\mathrm{bath}$.

We conclude that overall, all observables agree well with the benchmarks and prove the feasibility of our procedure over all considered parameter regimes in the SIAM model. Our results could be also improved for a given number of bath sites by switching from even to logarithmic spacing of bath energies in the model. This model-specific optimization will be the subject of a future study. The efficiency of the selected basis in the calculations can be used either to improve accuracy at given cluster sizes or to push the limits of treatable system sizes at a given accuracy. In the next section we will analyse the performance of our procedure in more detail.

\subsection{Results II: Performance of selective CI}
\label{section:results2}

\begin{figure}[t]
     \begin{center}
     \includegraphics[width=0.85\columnwidth]{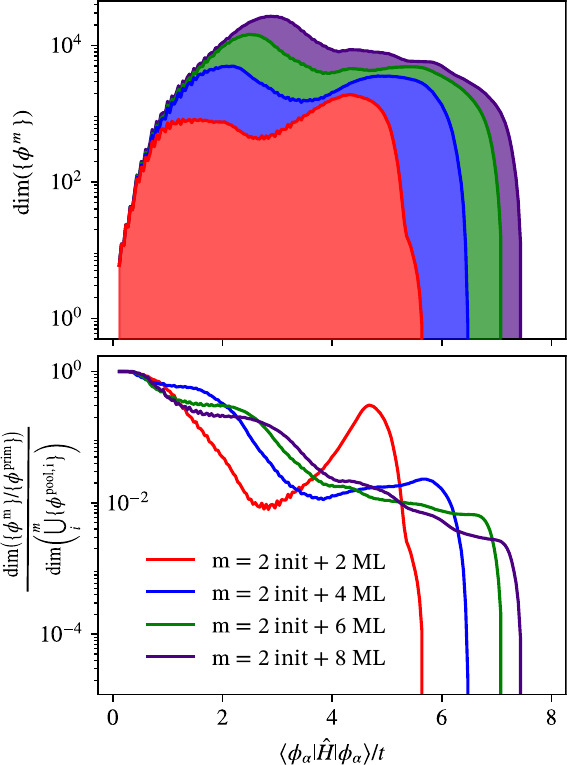}
     \end{center}
     \caption{Upper panel: distribution of selected determinants (cumulative) up to eight NN-supported iterations.
       Bottom panel: (cumulative) distribution of selected determinants normalized to the pool size of the corresponding iteration.
     }
     \label{fig:dist_fract}
\end{figure}

\begin{figure}[t]
     \begin{center}
     \includegraphics[width=0.45\textwidth]{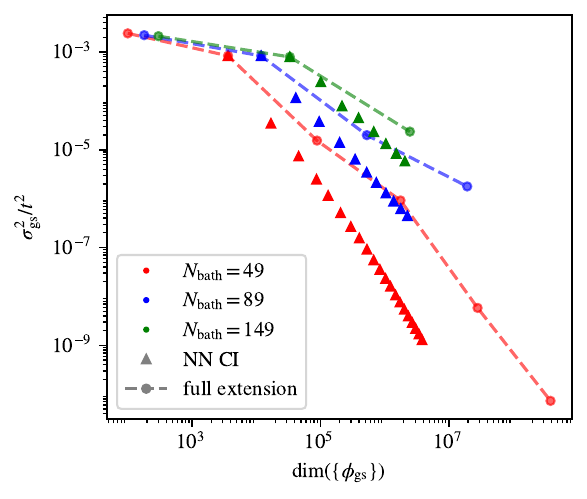}
     \end{center}
     \caption{Convergence of the variance of the ground state energy as a function of the basis dimension comparing NN selected bases (triangles) to bases by untruncated extensions
     (cicles) for three different system sizes.} 
     \label{fig:Energy_conv}
\end{figure}

We begin with a more thorough analysis of the basis sets constructed for the case with $N_\text{bath}=89$ which was discussed in the previous section. Starting with the spin symmetrized zero hybridization solution \eqref{eq:phi_init}, two initialisation extensions result in our first primary set of dimension $\text{dim}({\cal H}^{\text{primary}})= 11972$.

In the upper panel of Fig.~\ref{fig:alg} (b) we show the distribution of SDets in the first primary set in gray color as a function of their respective energy matrix element of the full Hamiltonian for an interaction strength of $U=4t$. In the same plot we further show the \emph{pool} (dark blue), the sampled \emph{training set} (light blue), the \emph{NN proposed} (orange), and finally the selected (red) determinants. While the identical shape of \emph{pool} and \emph{training set} reflects the random sampling strategy, the distribution of NN-proposed as well as finally selected determinants show a clear bias towards smaller energy around an energy of $4t$. As this energy approximately corresponds to the cost of an impurity double occupancy, the corresponding hole in the bath was created in a bath site with energy close to zero. In any case, even for the first NN selection the distribution is not at all well approximated by a naive energy cutoff due to an emerging fine-structure in the distribution which survives also after removing irrelevant determinants in the last step of our algorithm. The same conclusion can be drawn from plotting the fraction of proposed/selected determinants with respect to the pool, which is shown in the lower panel of Fig.~\ref{fig:alg} (b).
We continue our analysis by plotting the evolution of the energy distribution of the accumulated selected determinants over up to eight ML iterations (after $k=2$ initialization extensions) in Fig. \ref{fig:dist_fract}. The overall distribution and the selected pool fraction reveal that more subtle features start to emerge in the distribution. The larger peak for $m=2+2$ around an energy of $\approx4t\approx U$ can be understood as before by many configurations with a doubly occupied impurity sites and holes in "shallow" (i.e. $\varepsilon_b\approx0$) bath sites. For a larger number of extension steps the structure in the accumulated SDets and the respective selected fraction becomes more subtle.

Next, in Fig.~\ref{fig:Energy_conv} we plot the variance of the ground state energy as a function of the dimension of the corresponding Hilbert space (i.e. the number of contributing basis determinants). We compare the convergence of NN assisted iterations (triangles) with the behaviour of non-truncated extension 
steps. Starting from a small many-particle Hilbert space $\{\phi^\text{init}\}$, we expand the Hilbert space adding all determinants that couple to the original basis via the Hamiltonian. The comparison is done for three different model sizes with $N_{\text{bath}}\in\{49,89,149\}$. We find clear evidence that while the NN assisted scheme requires overall more iterations to reach a given accuracy, it significantly reduces the growth of the Hilbert space while systematically improving the accuracy of the ground state reaching variances of the order of $10^{-6}$ with a basis that is at least one order of magnitude smaller than the one after bare extensions.

Thus far we have quantified the effect of the NN selection step on the basis efficiency when compared to brute-force full extension steps. Next we can compare the NN CI deficiency to a selective CI scheme which relies on non-ML supported truncation steps after extensions such as implemented in \textsc{Quanty} \cite{Lu2014} which we will refer to as tCI in the following. The tCI approach also starts from a small many-body Hilbert space and extends the basis by acting with the Hamilton operator. Different from the NN CI approach, however, diagonalization in tCI is performed on the complete extended Hilbert space and SDets are selected/removed based on the norm square of their expansion coefficient and a cutoff parameter. Details of the iterative tCI scheme which was developed and implemented in \textsc{Quanty} can be found in \cite{Lu2014} and references therein.

\begin{figure}[t]
     \begin{center}
     \includegraphics[width=0.45\textwidth]{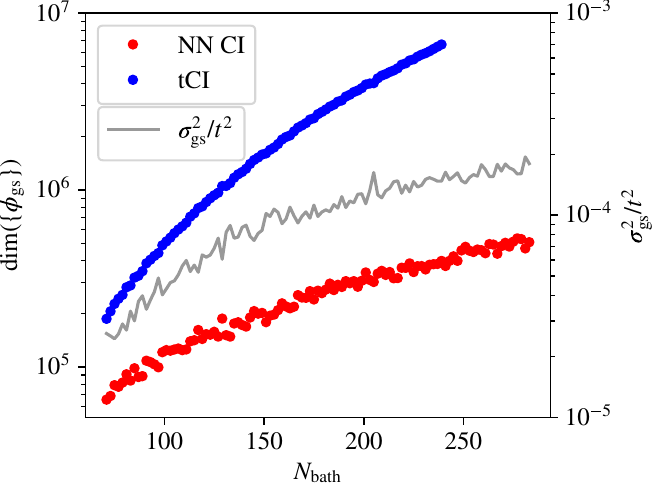}
     \end{center}
     \caption{Comparison of our NN CI results to a non ML truncation scheme as implemented in \cite{Lu2014}. Shown are the number of determinants (blue/red circles - left axis) required for a fixed variance (gray line - right axis). Red circles: simple truncation scheme (tCI); Blue circles: NN selection}
     \label{fig:benchmarkQuanty}
\end{figure}

In Fig.~\ref{fig:benchmarkQuanty}, we show the comparison of basis sizes reached with the two approaches (NN CI in red circles, tCI in blue circles) at equal variance (grey solid line). For this comparison, the NN CI calculation was fixed to two initialization steps ($k=2$) and two subsequent machine learning iterations - the tCI calculation was afterwards converged to the same variance. This restriction allowed us to reach system sizes up to $299$ bath sites. The results shown in Fig.~\ref{fig:benchmarkQuanty} demonstrate again an improved efficiency of the NN selected basis which i) for larger system sizes produces results on bases which are one order of magnitude smaller than the non ML scheme at equal accuracy, and ii) shows a significantly slower increase in the necessary basis sizes as a function of the system size. We note that for the considered case the tCI procedure removes just a small fraction of candidate SDets compared to a full extension step.

The results shown in Figs.~\ref{fig:Energy_conv} and \ref{fig:benchmarkQuanty} underline that we can exploit the NN selected bases both for increasing maximally feasibile accuracy or size of the model. Their consistency (each of the red circles presents an independent calculation) demonstrate that our algorithm yields reproducible results for variable model parameters. We close this section by mentioning once more (see end of section \ref{section:application}), that our scheme allows for various adjustments. The presented feasibility study thus opens the path towards application of our algorithm to a variety of quantum-cluster models in the context of solid state research and beyond.

\section{Summary and Conclusions}
\label{section:summary}
In summary, we propose a new algorithm for performing CI calculations, which aims to optimize the efficiency of the basis in high dimensional many-body Hilbert spaces. Our approach leverages a NN as  classifier for basis SDets within an iterative scheme. In the current work we have targeted computation of the ground state and its corresponding energy on a maximally efficient basis. By utilizing the NN in an active learning framework, the iterative algorithm actively selects and refines the subspace of determinants that significantly contribute to the ground state, thus reducing the computational burden while maintaining high accuracy.

The results from our study on the paradigmatic single impurity Anderson model confirm that the proposed NN-supported CI algorithm not only functions effectively, but also systematically and significantly outperforms other truncation schemes. The stability of our algorithm is evidenced by the reproducibility of results across independent calculations with different model and calculation parameters. The successful application of this method opens new avenues for tackling the 
``exponential wall" in full CI calculations, presenting practical implications for both solid-state physics models and exact calculations in quantum chemistry, e.g. for computing binding energies for small molecules complementary to existing ML supported approaches. In subsequent studies, we plan to explore the impact of different NN architectures and optimized convergence schemes on the performance of our algorithm, as well as the effects of various choices in the initial basis set, to further enhance the method's efficiency and applicability.

\begin{acknowledgements}
We thank F. Marquardt, P. Fadler, Y.~L.~A. Schmerwitz, G. Levi, E.~O.~Jonsson, and H. Jonsson for fruitful discussions. The authors gratefully acknowledge the scientific support and HPC resources provided by the Erlangen National High Performance Computing Center (NHR@FAU) of the Friedrich-Alexander-Universität Erlangen-Nürnberg (FAU). AP  acknowledges funding from the Deutsche Forschungsgemeinschaft (DFG) through the Heisenberg Program (Project PA2508/3-1) and the Cluster of Excellence on Complexity and Topology in Quantum Matter - ct.qmat (EXC 2147, Project No. 390858490).
\end{acknowledgements}
\bibliographystyle{apsrev4-1}
\bibliography{refs}

\begin{thebibliography}{53}%
\makeatletter
\providecommand \@ifxundefined [1]{%
 \@ifx{#1\undefined}
}%
\providecommand \@ifnum [1]{%
 \ifnum #1\expandafter \@firstoftwo
 \else \expandafter \@secondoftwo
 \fi
}%
\providecommand \@ifx [1]{%
 \ifx #1\expandafter \@firstoftwo
 \else \expandafter \@secondoftwo
 \fi
}%
\providecommand \natexlab [1]{#1}%
\providecommand \enquote  [1]{``#1''}%
\providecommand \bibnamefont  [1]{#1}%
\providecommand \bibfnamefont [1]{#1}%
\providecommand \citenamefont [1]{#1}%
\providecommand \href@noop [0]{\@secondoftwo}%
\providecommand \href [0]{\begingroup \@sanitize@url \@href}%
\providecommand \@href[1]{\@@startlink{#1}\@@href}%
\providecommand \@@href[1]{\endgroup#1\@@endlink}%
\providecommand \@sanitize@url [0]{\catcode `\\12\catcode `\$12\catcode
  `\&12\catcode `\#12\catcode `\^12\catcode `\_12\catcode `\%12\relax}%
\providecommand \@@startlink[1]{}%
\providecommand \@@endlink[0]{}%
\providecommand \url  [0]{\begingroup\@sanitize@url \@url }%
\providecommand \@url [1]{\endgroup\@href {#1}{\urlprefix }}%
\providecommand \urlprefix  [0]{URL }%
\providecommand \Eprint [0]{\href }%
\providecommand \doibase [0]{http://dx.doi.org/}%
\providecommand \selectlanguage [0]{\@gobble}%
\providecommand \bibinfo  [0]{\@secondoftwo}%
\providecommand \bibfield  [0]{\@secondoftwo}%
\providecommand \translation [1]{[#1]}%
\providecommand \BibitemOpen [0]{}%
\providecommand \bibitemStop [0]{}%
\providecommand \bibitemNoStop [0]{.\EOS\space}%
\providecommand \EOS [0]{\spacefactor3000\relax}%
\providecommand \BibitemShut  [1]{\csname bibitem#1\endcsname}%
\let\auto@bib@innerbib\@empty
\bibitem [{\citenamefont {Szabo}\ and\ \citenamefont
  {Ostlund}(2012)}]{Szabo_ModnQChem_2012}%
  \BibitemOpen
  \bibfield  {author} {\bibinfo {author} {\bibfnamefont {A.}~\bibnamefont
  {Szabo}}\ and\ \bibinfo {author} {\bibfnamefont {N.}~\bibnamefont
  {Ostlund}},\ }\href {https://books.google.de/books?id=KQ3DAgAAQBAJ} {\emph
  {\bibinfo {title} {Modern Quantum Chemistry: Introduction to Advanced
  Electronic Structure Theory}}},\ Dover Books on Chemistry\ (\bibinfo
  {publisher} {Dover Publications},\ \bibinfo {year} {2012})\BibitemShut
  {NoStop}%
\bibitem [{\citenamefont {Rossi}\ \emph {et~al.}(1999)\citenamefont {Rossi},
  \citenamefont {Bendazzoli}, \citenamefont {Evangelisti},\ and\ \citenamefont
  {Maynau}}]{ROSSI1999}%
  \BibitemOpen
  \bibfield  {author} {\bibinfo {author} {\bibfnamefont {E.}~\bibnamefont
  {Rossi}}, \bibinfo {author} {\bibfnamefont {G.~L.}\ \bibnamefont
  {Bendazzoli}}, \bibinfo {author} {\bibfnamefont {S.}~\bibnamefont
  {Evangelisti}}, \ and\ \bibinfo {author} {\bibfnamefont {D.}~\bibnamefont
  {Maynau}},\ }\href {\doibase https://doi.org/10.1016/S0009-2614(99)00791-5}
  {\bibfield  {journal} {\bibinfo  {journal} {Chemical Physics Letters}\
  }\textbf {\bibinfo {volume} {310}},\ \bibinfo {pages} {530} (\bibinfo {year}
  {1999})}\BibitemShut {NoStop}%
\bibitem [{\citenamefont {Roos}\ \emph {et~al.}(1980)\citenamefont {Roos},
  \citenamefont {Taylor},\ and\ \citenamefont {Sigbahn}}]{MCSCF01}%
  \BibitemOpen
  \bibfield  {author} {\bibinfo {author} {\bibfnamefont {B.~O.}\ \bibnamefont
  {Roos}}, \bibinfo {author} {\bibfnamefont {P.~R.}\ \bibnamefont {Taylor}}, \
  and\ \bibinfo {author} {\bibfnamefont {P.~E.}\ \bibnamefont {Sigbahn}},\
  }\href {\doibase https://doi.org/10.1016/0301-0104(80)80045-0} {\bibfield
  {journal} {\bibinfo  {journal} {Chemical Physics}\ }\textbf {\bibinfo
  {volume} {48}},\ \bibinfo {pages} {157} (\bibinfo {year} {1980})}\BibitemShut
  {NoStop}%
\bibitem [{\citenamefont {Werner}\ and\ \citenamefont
  {Knowles}(1985)}]{MCSCF02}%
  \BibitemOpen
  \bibfield  {author} {\bibinfo {author} {\bibfnamefont {H.}~\bibnamefont
  {Werner}}\ and\ \bibinfo {author} {\bibfnamefont {P.~J.}\ \bibnamefont
  {Knowles}},\ }\href {\doibase 10.1063/1.448627} {\bibfield  {journal}
  {\bibinfo  {journal} {The Journal of Chemical Physics}\ }\textbf {\bibinfo
  {volume} {82}},\ \bibinfo {pages} {5053} (\bibinfo {year} {1985})},\ \Eprint
  {http://arxiv.org/abs/https://pubs.aip.org/aip/jcp/article-pdf/82/11/5053/18953065/5053\_1\_online.pdf}
  {https://pubs.aip.org/aip/jcp/article-pdf/82/11/5053/18953065/5053\_1\_online.pdf}
  \BibitemShut {NoStop}%
\bibitem [{\citenamefont {{David Sherrill}}\ and\ \citenamefont
  {Schaefer}(1999)}]{MCSCF03}%
  \BibitemOpen
  \bibfield  {author} {\bibinfo {author} {\bibfnamefont {C.}~\bibnamefont
  {{David Sherrill}}}\ and\ \bibinfo {author} {\bibfnamefont {H.~F.}\
  \bibnamefont {Schaefer}}\ }(\bibinfo  {publisher} {Academic Press},\ \bibinfo
  {year} {1999})\ pp.\ \bibinfo {pages} {143--269}\BibitemShut {NoStop}%
\bibitem [{\citenamefont {Metzner}\ and\ \citenamefont
  {Vollhardt}(1989)}]{dmft1}%
  \BibitemOpen
  \bibfield  {author} {\bibinfo {author} {\bibfnamefont {W.}~\bibnamefont
  {Metzner}}\ and\ \bibinfo {author} {\bibfnamefont {D.}~\bibnamefont
  {Vollhardt}},\ }\href {\doibase 10.1103/PhysRevLett.62.324} {\bibfield
  {journal} {\bibinfo  {journal} {Phys. Rev. Lett.}\ }\textbf {\bibinfo
  {volume} {62}},\ \bibinfo {pages} {324} (\bibinfo {year} {1989})}\BibitemShut
  {NoStop}%
\bibitem [{\citenamefont {Georges}\ and\ \citenamefont
  {Kotliar}(1992)}]{Georges1992}%
  \BibitemOpen
  \bibfield  {author} {\bibinfo {author} {\bibfnamefont {A.}~\bibnamefont
  {Georges}}\ and\ \bibinfo {author} {\bibfnamefont {G.}~\bibnamefont
  {Kotliar}},\ }\href {\doibase 10.1103/PhysRevB.45.6479} {\bibfield  {journal}
  {\bibinfo  {journal} {Phys. Rev. B}\ }\textbf {\bibinfo {volume} {45}},\
  \bibinfo {pages} {6479} (\bibinfo {year} {1992})}\BibitemShut {NoStop}%
\bibitem [{\citenamefont {Georges}\ \emph {et~al.}(1996)\citenamefont
  {Georges}, \citenamefont {Kotliar}, \citenamefont {Krauth},\ and\
  \citenamefont {Rozenberg}}]{dmft2}%
  \BibitemOpen
  \bibfield  {author} {\bibinfo {author} {\bibfnamefont {A.}~\bibnamefont
  {Georges}}, \bibinfo {author} {\bibfnamefont {G.}~\bibnamefont {Kotliar}},
  \bibinfo {author} {\bibfnamefont {W.}~\bibnamefont {Krauth}}, \ and\ \bibinfo
  {author} {\bibfnamefont {M.~J.}\ \bibnamefont {Rozenberg}},\ }\href {\doibase
  10.1103/RevModPhys.68.13} {\bibfield  {journal} {\bibinfo  {journal} {Rev.
  Mod. Phys.}\ }\textbf {\bibinfo {volume} {68}},\ \bibinfo {pages} {13}
  (\bibinfo {year} {1996})}\BibitemShut {NoStop}%
\bibitem [{\citenamefont {Lichtenstein}\ and\ \citenamefont
  {Katsnelson}(2000)}]{cdmft1}%
  \BibitemOpen
  \bibfield  {author} {\bibinfo {author} {\bibfnamefont {A.~I.}\ \bibnamefont
  {Lichtenstein}}\ and\ \bibinfo {author} {\bibfnamefont {M.~I.}\ \bibnamefont
  {Katsnelson}},\ }\href {\doibase 10.1103/PhysRevB.62.R9283} {\bibfield
  {journal} {\bibinfo  {journal} {Phys. Rev. B}\ }\textbf {\bibinfo {volume}
  {62}},\ \bibinfo {pages} {R9283} (\bibinfo {year} {2000})}\BibitemShut
  {NoStop}%
\bibitem [{\citenamefont {Kotliar}\ \emph {et~al.}(2001)\citenamefont
  {Kotliar}, \citenamefont {Savrasov}, \citenamefont {P\'alsson},\ and\
  \citenamefont {Biroli}}]{cdmft2}%
  \BibitemOpen
  \bibfield  {author} {\bibinfo {author} {\bibfnamefont {G.}~\bibnamefont
  {Kotliar}}, \bibinfo {author} {\bibfnamefont {S.~Y.}\ \bibnamefont
  {Savrasov}}, \bibinfo {author} {\bibfnamefont {G.}~\bibnamefont {P\'alsson}},
  \ and\ \bibinfo {author} {\bibfnamefont {G.}~\bibnamefont {Biroli}},\ }\href
  {\doibase 10.1103/PhysRevLett.87.186401} {\bibfield  {journal} {\bibinfo
  {journal} {Phys. Rev. Lett.}\ }\textbf {\bibinfo {volume} {87}},\ \bibinfo
  {pages} {186401} (\bibinfo {year} {2001})}\BibitemShut {NoStop}%
\bibitem [{\citenamefont {Ivanic}\ and\ \citenamefont
  {Ruedenberg}(2001)}]{Ivanic2001}%
  \BibitemOpen
  \bibfield  {author} {\bibinfo {author} {\bibfnamefont {J.}~\bibnamefont
  {Ivanic}}\ and\ \bibinfo {author} {\bibfnamefont {K.}~\bibnamefont
  {Ruedenberg}},\ }\href {\doibase https://doi.org/10.1007/s002140100285}
  {\bibfield  {journal} {\bibinfo  {journal} {Theor. Chem. Acc.}\ }\textbf
  {\bibinfo {volume} {106}},\ \bibinfo {pages} {339} (\bibinfo {year}
  {2001})}\BibitemShut {NoStop}%
\bibitem [{\citenamefont {Huron}\ \emph {et~al.}(1973)\citenamefont {Huron},
  \citenamefont {Malrieu},\ and\ \citenamefont {Rancurel}}]{Huron1973}%
  \BibitemOpen
  \bibfield  {author} {\bibinfo {author} {\bibfnamefont {B.}~\bibnamefont
  {Huron}}, \bibinfo {author} {\bibfnamefont {J.~P.}\ \bibnamefont {Malrieu}},
  \ and\ \bibinfo {author} {\bibfnamefont {P.}~\bibnamefont {Rancurel}},\
  }\href {\doibase 10.1063/1.1679199} {\bibfield  {journal} {\bibinfo
  {journal} {The Journal of Chemical Physics}\ }\textbf {\bibinfo {volume}
  {58}},\ \bibinfo {pages} {5745} (\bibinfo {year} {1973})},\ \Eprint
  {http://arxiv.org/abs/https://pubs.aip.org/aip/jcp/article-pdf/58/12/5745/18885418/5745\_1\_online.pdf}
  {https://pubs.aip.org/aip/jcp/article-pdf/58/12/5745/18885418/5745\_1\_online.pdf}
  \BibitemShut {NoStop}%
\bibitem [{\citenamefont {Greer}(1998)}]{Greer1998}%
  \BibitemOpen
  \bibfield  {author} {\bibinfo {author} {\bibfnamefont {J.}~\bibnamefont
  {Greer}},\ }\href {\doibase https://doi.org/10.1006/jcph.1998.5953}
  {\bibfield  {journal} {\bibinfo  {journal} {Journal of Computational
  Physics}\ }\textbf {\bibinfo {volume} {146}},\ \bibinfo {pages} {181}
  (\bibinfo {year} {1998})}\BibitemShut {NoStop}%
\bibitem [{\citenamefont {Garniron}\ \emph {et~al.}(2018)\citenamefont
  {Garniron}, \citenamefont {Scemama}, \citenamefont {Giner}, \citenamefont
  {Caffarel},\ and\ \citenamefont {Loos}}]{Garniron2018}%
  \BibitemOpen
  \bibfield  {author} {\bibinfo {author} {\bibfnamefont {Y.}~\bibnamefont
  {Garniron}}, \bibinfo {author} {\bibfnamefont {A.}~\bibnamefont {Scemama}},
  \bibinfo {author} {\bibfnamefont {E.}~\bibnamefont {Giner}}, \bibinfo
  {author} {\bibfnamefont {M.}~\bibnamefont {Caffarel}}, \ and\ \bibinfo
  {author} {\bibfnamefont {P.-F.}\ \bibnamefont {Loos}},\ }\href {\doibase
  10.1063/1.5044503} {\bibfield  {journal} {\bibinfo  {journal} {The Journal of
  Chemical Physics}\ }\textbf {\bibinfo {volume} {149}},\ \bibinfo {pages}
  {064103} (\bibinfo {year} {2018})},\ \Eprint
  {http://arxiv.org/abs/https://pubs.aip.org/aip/jcp/article-pdf/doi/10.1063/1.5044503/13501037/064103\_1\_online.pdf}
  {https://pubs.aip.org/aip/jcp/article-pdf/doi/10.1063/1.5044503/13501037/064103\_1\_online.pdf}
  \BibitemShut {NoStop}%
\bibitem [{\citenamefont {Tubman}\ \emph {et~al.}(2020)\citenamefont {Tubman},
  \citenamefont {Freeman}, \citenamefont {Levine}, \citenamefont {Hait},
  \citenamefont {Head-Gordon},\ and\ \citenamefont {Whaley}}]{Tubman2020}%
  \BibitemOpen
  \bibfield  {author} {\bibinfo {author} {\bibfnamefont {N.~M.}\ \bibnamefont
  {Tubman}}, \bibinfo {author} {\bibfnamefont {C.~D.}\ \bibnamefont {Freeman}},
  \bibinfo {author} {\bibfnamefont {D.~S.}\ \bibnamefont {Levine}}, \bibinfo
  {author} {\bibfnamefont {D.}~\bibnamefont {Hait}}, \bibinfo {author}
  {\bibfnamefont {M.}~\bibnamefont {Head-Gordon}}, \ and\ \bibinfo {author}
  {\bibfnamefont {K.~B.}\ \bibnamefont {Whaley}},\ }\href {\doibase
  10.1021/acs.jctc.8b00536} {\bibfield  {journal} {\bibinfo  {journal} {Journal
  of Chemical Theory and Computation}\ }\textbf {\bibinfo {volume} {16}},\
  \bibinfo {pages} {2139} (\bibinfo {year} {2020})},\ \bibinfo {note} {pMID:
  32159951},\ \Eprint
  {http://arxiv.org/abs/https://doi.org/10.1021/acs.jctc.8b00536}
  {https://doi.org/10.1021/acs.jctc.8b00536} \BibitemShut {NoStop}%
\bibitem [{\citenamefont {Coe}(2018)}]{Coe_MLCI_JChemTC_2018}%
  \BibitemOpen
  \bibfield  {author} {\bibinfo {author} {\bibfnamefont {J.~P.}\ \bibnamefont
  {Coe}},\ }\href {\doibase 10.1021/acs.jctc.8b00849} {\bibfield  {journal}
  {\bibinfo  {journal} {J. Chem. Theory Comput.}\ }\textbf {\bibinfo {volume}
  {14}},\ \bibinfo {pages} {5739} (\bibinfo {year} {2018})},\ \bibinfo {note}
  {pMID: 30285426},\ \Eprint
  {http://arxiv.org/abs/https://doi.org/10.1021/acs.jctc.8b00849}
  {https://doi.org/10.1021/acs.jctc.8b00849} \BibitemShut {NoStop}%
\bibitem [{\citenamefont {Jeong}\ \emph {et~al.}(2021)\citenamefont {Jeong},
  \citenamefont {Gaggioli},\ and\ \citenamefont
  {Gagliardi}}]{Jeong_ALCI_JChemTC_2021}%
  \BibitemOpen
  \bibfield  {author} {\bibinfo {author} {\bibfnamefont {W.}~\bibnamefont
  {Jeong}}, \bibinfo {author} {\bibfnamefont {C.~A.}\ \bibnamefont {Gaggioli}},
  \ and\ \bibinfo {author} {\bibfnamefont {L.}~\bibnamefont {Gagliardi}},\
  }\href {\doibase 10.1021/acs.jctc.1c00769} {\bibfield  {journal} {\bibinfo
  {journal} {Journal of Chemical Theory and Computation}\ }\textbf {\bibinfo
  {volume} {17}},\ \bibinfo {pages} {7518} (\bibinfo {year}
  {2021})}\BibitemShut {NoStop}%
\bibitem [{\citenamefont {Pineda~Flores}(2021)}]{Chembot}%
  \BibitemOpen
  \bibfield  {author} {\bibinfo {author} {\bibfnamefont {S.~D.}\ \bibnamefont
  {Pineda~Flores}},\ }\href {\doibase 10.1021/acs.jctc.1c00196} {\bibfield
  {journal} {\bibinfo  {journal} {Journal of Chemical Theory and Computation}\
  }\textbf {\bibinfo {volume} {17}},\ \bibinfo {pages} {4028} (\bibinfo {year}
  {2021})},\ \bibinfo {note} {pMID: 34125549},\ \Eprint
  {http://arxiv.org/abs/https://doi.org/10.1021/acs.jctc.1c00196}
  {https://doi.org/10.1021/acs.jctc.1c00196} \BibitemShut {NoStop}%
\bibitem [{\citenamefont {Goings}\ \emph {et~al.}(2021)\citenamefont {Goings},
  \citenamefont {Hu}, \citenamefont {Yang},\ and\ \citenamefont {Li}}]{RLCI}%
  \BibitemOpen
  \bibfield  {author} {\bibinfo {author} {\bibfnamefont {J.~J.}\ \bibnamefont
  {Goings}}, \bibinfo {author} {\bibfnamefont {H.}~\bibnamefont {Hu}}, \bibinfo
  {author} {\bibfnamefont {C.}~\bibnamefont {Yang}}, \ and\ \bibinfo {author}
  {\bibfnamefont {X.}~\bibnamefont {Li}},\ }\href {\doibase
  10.1021/acs.jctc.1c00010} {\bibfield  {journal} {\bibinfo  {journal} {Journal
  of Chemical Theory and Computation}\ }\textbf {\bibinfo {volume} {17}},\
  \bibinfo {pages} {5482} (\bibinfo {year} {2021})},\ \bibinfo {note} {pMID:
  34423637},\ \Eprint
  {http://arxiv.org/abs/https://doi.org/10.1021/acs.jctc.1c00010}
  {https://doi.org/10.1021/acs.jctc.1c00010} \BibitemShut {NoStop}%
\bibitem [{\citenamefont {Herzog}\ \emph {et~al.}(2023)\citenamefont {Herzog},
  \citenamefont {Casier}, \citenamefont {Lebegue},\ and\ \citenamefont
  {Rocca}}]{Herzog2023}%
  \BibitemOpen
  \bibfield  {author} {\bibinfo {author} {\bibfnamefont {B.}~\bibnamefont
  {Herzog}}, \bibinfo {author} {\bibfnamefont {B.}~\bibnamefont {Casier}},
  \bibinfo {author} {\bibfnamefont {S.}~\bibnamefont {Lebegue}}, \ and\
  \bibinfo {author} {\bibfnamefont {D.}~\bibnamefont {Rocca}},\ }\href
  {\doibase 10.1021/acs.jctc.2c01216} {\bibfield  {journal} {\bibinfo
  {journal} {Journal of Chemical Theory and Computation}\ }\textbf {\bibinfo
  {volume} {19}},\ \bibinfo {pages} {2484} (\bibinfo {year} {2023})},\ \Eprint
  {http://arxiv.org/abs/https://doi.org/10.1021/acs.jctc.2c01216}
  {https://doi.org/10.1021/acs.jctc.2c01216} \BibitemShut {NoStop}%
\bibitem [{\citenamefont {Coe}(2019)}]{Coe_JChemTC_2019}%
  \BibitemOpen
  \bibfield  {author} {\bibinfo {author} {\bibfnamefont {J.~P.}\ \bibnamefont
  {Coe}},\ }\href {\doibase https://doi.org/10.1021/acs.jctc.9b00828}
  {\bibfield  {journal} {\bibinfo  {journal} {J. Chem. Theory Comput.}\
  }\textbf {\bibinfo {volume} {15}},\ \bibinfo {pages} {6179} (\bibinfo {year}
  {2019})}\BibitemShut {NoStop}%
\bibitem [{\citenamefont {Molchanov}\ \emph {et~al.}(2022)\citenamefont
  {Molchanov}, \citenamefont {Launey}, \citenamefont {Mercenne}, \citenamefont
  {Sargsyan}, \citenamefont {Dytrych},\ and\ \citenamefont
  {Draayer}}]{Molchanov2022}%
  \BibitemOpen
  \bibfield  {author} {\bibinfo {author} {\bibfnamefont {O.~M.}\ \bibnamefont
  {Molchanov}}, \bibinfo {author} {\bibfnamefont {K.~D.}\ \bibnamefont
  {Launey}}, \bibinfo {author} {\bibfnamefont {A.}~\bibnamefont {Mercenne}},
  \bibinfo {author} {\bibfnamefont {G.~H.}\ \bibnamefont {Sargsyan}}, \bibinfo
  {author} {\bibfnamefont {T.}~\bibnamefont {Dytrych}}, \ and\ \bibinfo
  {author} {\bibfnamefont {J.~P.}\ \bibnamefont {Draayer}},\ }\href {\doibase
  10.1103/PhysRevC.105.034306} {\bibfield  {journal} {\bibinfo  {journal}
  {Phys. Rev. C}\ }\textbf {\bibinfo {volume} {105}},\ \bibinfo {pages}
  {034306} (\bibinfo {year} {2022})}\BibitemShut {NoStop}%
\bibitem [{\citenamefont {Bilous}\ \emph {et~al.}(2023)\citenamefont {Bilous},
  \citenamefont {P\'alffy},\ and\ \citenamefont {Marquardt}}]{MLGRASP}%
  \BibitemOpen
  \bibfield  {author} {\bibinfo {author} {\bibfnamefont {P.}~\bibnamefont
  {Bilous}}, \bibinfo {author} {\bibfnamefont {A.}~\bibnamefont {P\'alffy}}, \
  and\ \bibinfo {author} {\bibfnamefont {F.}~\bibnamefont {Marquardt}},\ }\href
  {\doibase 10.1103/PhysRevLett.131.133002} {\bibfield  {journal} {\bibinfo
  {journal} {Phys. Rev. Lett.}\ }\textbf {\bibinfo {volume} {131}},\ \bibinfo
  {pages} {133002} (\bibinfo {year} {2023})}\BibitemShut {NoStop}%
\bibitem [{\citenamefont {Anderson}(1961)}]{And1961}%
  \BibitemOpen
  \bibfield  {author} {\bibinfo {author} {\bibfnamefont {P.~W.}\ \bibnamefont
  {Anderson}},\ }\href {\doibase 10.1103/PhysRev.124.41} {\bibfield  {journal}
  {\bibinfo  {journal} {Phys. Rev.}\ }\textbf {\bibinfo {volume} {124}},\
  \bibinfo {pages} {41} (\bibinfo {year} {1961})}\BibitemShut {NoStop}%
\bibitem [{\citenamefont {Nuss}\ \emph {et~al.}(2012)\citenamefont {Nuss},
  \citenamefont {Arrigoni}, \citenamefont {Aichhorn},\ and\ \citenamefont
  {von~der Linden}}]{Aichhorn2011}%
  \BibitemOpen
  \bibfield  {author} {\bibinfo {author} {\bibfnamefont {M.}~\bibnamefont
  {Nuss}}, \bibinfo {author} {\bibfnamefont {E.}~\bibnamefont {Arrigoni}},
  \bibinfo {author} {\bibfnamefont {M.}~\bibnamefont {Aichhorn}}, \ and\
  \bibinfo {author} {\bibfnamefont {W.}~\bibnamefont {von~der Linden}},\ }\href
  {\doibase 10.1103/PhysRevB.85.235107} {\bibfield  {journal} {\bibinfo
  {journal} {Phys. Rev. B}\ }\textbf {\bibinfo {volume} {85}},\ \bibinfo
  {pages} {235107} (\bibinfo {year} {2012})}\BibitemShut {NoStop}%
\bibitem [{\citenamefont {Lu}\ \emph {et~al.}(2014)\citenamefont {Lu},
  \citenamefont {H\"oppner}, \citenamefont {Gunnarsson},\ and\ \citenamefont
  {Haverkort}}]{Lu2014}%
  \BibitemOpen
  \bibfield  {author} {\bibinfo {author} {\bibfnamefont {Y.}~\bibnamefont
  {Lu}}, \bibinfo {author} {\bibfnamefont {M.}~\bibnamefont {H\"oppner}},
  \bibinfo {author} {\bibfnamefont {O.}~\bibnamefont {Gunnarsson}}, \ and\
  \bibinfo {author} {\bibfnamefont {M.~W.}\ \bibnamefont {Haverkort}},\ }\href
  {\doibase 10.1103/PhysRevB.90.085102} {\bibfield  {journal} {\bibinfo
  {journal} {Phys. Rev. B}\ }\textbf {\bibinfo {volume} {90}},\ \bibinfo
  {pages} {085102} (\bibinfo {year} {2014})}\BibitemShut {NoStop}%
\bibitem [{\citenamefont {Cao}\ \emph {et~al.}(2021)\citenamefont {Cao},
  \citenamefont {Lu}, \citenamefont {Hansmann},\ and\ \citenamefont
  {Haverkort}}]{Cao2021}%
  \BibitemOpen
  \bibfield  {author} {\bibinfo {author} {\bibfnamefont {X.}~\bibnamefont
  {Cao}}, \bibinfo {author} {\bibfnamefont {Y.}~\bibnamefont {Lu}}, \bibinfo
  {author} {\bibfnamefont {P.}~\bibnamefont {Hansmann}}, \ and\ \bibinfo
  {author} {\bibfnamefont {M.~W.}\ \bibnamefont {Haverkort}},\ }\href {\doibase
  10.1103/PhysRevB.104.115119} {\bibfield  {journal} {\bibinfo  {journal}
  {Phys. Rev. B}\ }\textbf {\bibinfo {volume} {104}},\ \bibinfo {pages}
  {115119} (\bibinfo {year} {2021})}\BibitemShut {NoStop}%
\bibitem [{\citenamefont {Carleo}\ \emph {et~al.}(2019)\citenamefont {Carleo},
  \citenamefont {Cirac}, \citenamefont {Cranmer}, \citenamefont {Daudet},
  \citenamefont {Schuld}, \citenamefont {Tishby}, \citenamefont
  {Vogt-Maranto},\ and\ \citenamefont {Zdeborov\'a}}]{Carleo2019}%
  \BibitemOpen
  \bibfield  {author} {\bibinfo {author} {\bibfnamefont {G.}~\bibnamefont
  {Carleo}}, \bibinfo {author} {\bibfnamefont {I.}~\bibnamefont {Cirac}},
  \bibinfo {author} {\bibfnamefont {K.}~\bibnamefont {Cranmer}}, \bibinfo
  {author} {\bibfnamefont {L.}~\bibnamefont {Daudet}}, \bibinfo {author}
  {\bibfnamefont {M.}~\bibnamefont {Schuld}}, \bibinfo {author} {\bibfnamefont
  {N.}~\bibnamefont {Tishby}}, \bibinfo {author} {\bibfnamefont
  {L.}~\bibnamefont {Vogt-Maranto}}, \ and\ \bibinfo {author} {\bibfnamefont
  {L.}~\bibnamefont {Zdeborov\'a}},\ }\href {\doibase
  10.1103/RevModPhys.91.045002} {\bibfield  {journal} {\bibinfo  {journal}
  {Rev. Mod. Phys.}\ }\textbf {\bibinfo {volume} {91}},\ \bibinfo {pages}
  {045002} (\bibinfo {year} {2019})}\BibitemShut {NoStop}%
\bibitem [{\citenamefont {Dey}\ and\ \citenamefont {Ghosh}(2023)}]{Dey2023}%
  \BibitemOpen
  \bibfield  {author} {\bibinfo {author} {\bibfnamefont {M.}~\bibnamefont
  {Dey}}\ and\ \bibinfo {author} {\bibfnamefont {D.}~\bibnamefont {Ghosh}},\
  }\href {\doibase 10.1021/acs.jpca.3c05322} {\bibfield  {journal} {\bibinfo
  {journal} {J.~Chem.~Phys.~A}\ }\textbf {\bibinfo {volume} {127}},\ \bibinfo
  {pages} {9159} (\bibinfo {year} {2023})},\ \Eprint
  {http://arxiv.org/abs/https://doi.org/10.1021/acs.jpca.3c05322}
  {https://doi.org/10.1021/acs.jpca.3c05322} \BibitemShut {NoStop}%
\bibitem [{\citenamefont {Carleo}\ and\ \citenamefont
  {Troyer}(2017)}]{Carleo2017}%
  \BibitemOpen
  \bibfield  {author} {\bibinfo {author} {\bibfnamefont {G.}~\bibnamefont
  {Carleo}}\ and\ \bibinfo {author} {\bibfnamefont {M.}~\bibnamefont
  {Troyer}},\ }\href {\doibase 10.1126/science.aag2302} {\bibfield  {journal}
  {\bibinfo  {journal} {Science}\ }\textbf {\bibinfo {volume} {355}},\ \bibinfo
  {pages} {602} (\bibinfo {year} {2017})},\ \Eprint
  {http://arxiv.org/abs/https://www.science.org/doi/pdf/10.1126/science.aag2302}
  {https://www.science.org/doi/pdf/10.1126/science.aag2302} \BibitemShut
  {NoStop}%
\bibitem [{\citenamefont {Schmitt}\ \emph {et~al.}(2022)\citenamefont
  {Schmitt}, \citenamefont {Rams}, \citenamefont {Dziarmaga}, \citenamefont
  {Heyl},\ and\ \citenamefont {Zurek}}]{SchmittRamsDziarmagaetal.2022}%
  \BibitemOpen
  \bibfield  {author} {\bibinfo {author} {\bibfnamefont {M.}~\bibnamefont
  {Schmitt}}, \bibinfo {author} {\bibfnamefont {M.~M.}\ \bibnamefont {Rams}},
  \bibinfo {author} {\bibfnamefont {J.}~\bibnamefont {Dziarmaga}}, \bibinfo
  {author} {\bibfnamefont {M.}~\bibnamefont {Heyl}}, \ and\ \bibinfo {author}
  {\bibfnamefont {W.~H.}\ \bibnamefont {Zurek}},\ }\href {\doibase
  10.1126/sciadv.abl6850} {\bibfield  {journal} {\bibinfo  {journal} {Science
  Advances}\ }\textbf {\bibinfo {volume} {8}},\ \bibinfo {pages} {eabl6850}
  (\bibinfo {year} {2022})}\BibitemShut {NoStop}%
\bibitem [{\citenamefont {Saito}(2017)}]{Saito2017}%
  \BibitemOpen
  \bibfield  {author} {\bibinfo {author} {\bibfnamefont {H.}~\bibnamefont
  {Saito}},\ }\href {\doibase 10.7566/JPSJ.86.093001} {\bibfield  {journal}
  {\bibinfo  {journal} {Journal of the Physical Society of Japan}\ }\textbf
  {\bibinfo {volume} {86}},\ \bibinfo {pages} {093001} (\bibinfo {year}
  {2017})},\ \Eprint
  {http://arxiv.org/abs/https://doi.org/10.7566/JPSJ.86.093001}
  {https://doi.org/10.7566/JPSJ.86.093001} \BibitemShut {NoStop}%
\bibitem [{\citenamefont {Choo}\ \emph {et~al.}(2020)\citenamefont {Choo},
  \citenamefont {Mezzacapo},\ and\ \citenamefont {Carleo}}]{Choo2020}%
  \BibitemOpen
  \bibfield  {author} {\bibinfo {author} {\bibfnamefont {K.}~\bibnamefont
  {Choo}}, \bibinfo {author} {\bibfnamefont {A.}~\bibnamefont {Mezzacapo}}, \
  and\ \bibinfo {author} {\bibfnamefont {G.}~\bibnamefont {Carleo}},\ }\href
  {\doibase 10.1038/s41467-020-15724-9} {\bibfield  {journal} {\bibinfo
  {journal} {Nature Communications}\ }\textbf {\bibinfo {volume} {11}},\
  \bibinfo {pages} {2368} (\bibinfo {year} {2020})}\BibitemShut {NoStop}%
\bibitem [{\citenamefont {Snyder}\ \emph {et~al.}(2012)\citenamefont {Snyder},
  \citenamefont {Rupp}, \citenamefont {Hansen}, \citenamefont {M\"uller},\ and\
  \citenamefont {Burke}}]{Snyder2012}%
  \BibitemOpen
  \bibfield  {author} {\bibinfo {author} {\bibfnamefont {J.~C.}\ \bibnamefont
  {Snyder}}, \bibinfo {author} {\bibfnamefont {M.}~\bibnamefont {Rupp}},
  \bibinfo {author} {\bibfnamefont {K.}~\bibnamefont {Hansen}}, \bibinfo
  {author} {\bibfnamefont {K.-R.}\ \bibnamefont {M\"uller}}, \ and\ \bibinfo
  {author} {\bibfnamefont {K.}~\bibnamefont {Burke}},\ }\href {\doibase
  10.1103/PhysRevLett.108.253002} {\bibfield  {journal} {\bibinfo  {journal}
  {Phys. Rev. Lett.}\ }\textbf {\bibinfo {volume} {108}},\ \bibinfo {pages}
  {253002} (\bibinfo {year} {2012})}\BibitemShut {NoStop}%
\bibitem [{\citenamefont {Seino}\ \emph {et~al.}(2018)\citenamefont {Seino},
  \citenamefont {Kageyama}, \citenamefont {Fujinami}, \citenamefont {Ikabata},\
  and\ \citenamefont {Nakai}}]{Seino2018}%
  \BibitemOpen
  \bibfield  {author} {\bibinfo {author} {\bibfnamefont {J.}~\bibnamefont
  {Seino}}, \bibinfo {author} {\bibfnamefont {R.}~\bibnamefont {Kageyama}},
  \bibinfo {author} {\bibfnamefont {M.}~\bibnamefont {Fujinami}}, \bibinfo
  {author} {\bibfnamefont {Y.}~\bibnamefont {Ikabata}}, \ and\ \bibinfo
  {author} {\bibfnamefont {H.}~\bibnamefont {Nakai}},\ }\href {\doibase
  10.1063/1.5007230} {\bibfield  {journal} {\bibinfo  {journal} {The Journal of
  Chemical Physics}\ }\textbf {\bibinfo {volume} {148}},\ \bibinfo {pages}
  {241705} (\bibinfo {year} {2018})},\ \Eprint
  {http://arxiv.org/abs/https://pubs.aip.org/aip/jcp/article-pdf/doi/10.1063/1.5007230/16652841/241705\_1\_online.pdf}
  {https://pubs.aip.org/aip/jcp/article-pdf/doi/10.1063/1.5007230/16652841/241705\_1\_online.pdf}
  \BibitemShut {NoStop}%
\bibitem [{\citenamefont {Dick}\ and\ \citenamefont
  {Fernandez-Serra}(2020)}]{Dick2020}%
  \BibitemOpen
  \bibfield  {author} {\bibinfo {author} {\bibfnamefont {S.}~\bibnamefont
  {Dick}}\ and\ \bibinfo {author} {\bibfnamefont {M.}~\bibnamefont
  {Fernandez-Serra}},\ }\href {\doibase 10.1038/s41467-020-17265-7} {\bibfield
  {journal} {\bibinfo  {journal} {Nature Communications}\ }\textbf {\bibinfo
  {volume} {11}},\ \bibinfo {pages} {3509} (\bibinfo {year}
  {2020})}\BibitemShut {NoStop}%
\bibitem [{\citenamefont {Brockherde}\ \emph {et~al.}(2017)\citenamefont
  {Brockherde}, \citenamefont {Vogt}, \citenamefont {Li}, \citenamefont
  {Tuckerman}, \citenamefont {Burke},\ and\ \citenamefont
  {M{\"u}ller}}]{Brockherde2017}%
  \BibitemOpen
  \bibfield  {author} {\bibinfo {author} {\bibfnamefont {F.}~\bibnamefont
  {Brockherde}}, \bibinfo {author} {\bibfnamefont {L.}~\bibnamefont {Vogt}},
  \bibinfo {author} {\bibfnamefont {L.}~\bibnamefont {Li}}, \bibinfo {author}
  {\bibfnamefont {M.~E.}\ \bibnamefont {Tuckerman}}, \bibinfo {author}
  {\bibfnamefont {K.}~\bibnamefont {Burke}}, \ and\ \bibinfo {author}
  {\bibfnamefont {K.-R.}\ \bibnamefont {M{\"u}ller}},\ }\href {\doibase
  10.1038/s41467-017-00839-3} {\bibfield  {journal} {\bibinfo  {journal}
  {Nature Communications}\ }\textbf {\bibinfo {volume} {8}},\ \bibinfo {pages}
  {872} (\bibinfo {year} {2017})}\BibitemShut {NoStop}%
\bibitem [{\citenamefont {Moreno}\ \emph {et~al.}(2020)\citenamefont {Moreno},
  \citenamefont {Carleo},\ and\ \citenamefont {Georges}}]{Moreno2020}%
  \BibitemOpen
  \bibfield  {author} {\bibinfo {author} {\bibfnamefont {J.~R.}\ \bibnamefont
  {Moreno}}, \bibinfo {author} {\bibfnamefont {G.}~\bibnamefont {Carleo}}, \
  and\ \bibinfo {author} {\bibfnamefont {A.}~\bibnamefont {Georges}},\ }\href
  {\doibase 10.1103/PhysRevLett.125.076402} {\bibfield  {journal} {\bibinfo
  {journal} {Phys. Rev. Lett.}\ }\textbf {\bibinfo {volume} {125}},\ \bibinfo
  {pages} {076402} (\bibinfo {year} {2020})}\BibitemShut {NoStop}%
\bibitem [{\citenamefont {del Rio}\ \emph {et~al.}(2023)\citenamefont {del
  Rio}, \citenamefont {Phan},\ and\ \citenamefont {Ramprasad}}]{DFTpseudoml1}%
  \BibitemOpen
  \bibfield  {author} {\bibinfo {author} {\bibfnamefont {B.~G.}\ \bibnamefont
  {del Rio}}, \bibinfo {author} {\bibfnamefont {B.}~\bibnamefont {Phan}}, \
  and\ \bibinfo {author} {\bibfnamefont {R.}~\bibnamefont {Ramprasad}},\ }\href
  {\doibase 10.1038/s41524-023-01115-3} {\bibfield  {journal} {\bibinfo
  {journal} {npj Computational Materials}\ }\textbf {\bibinfo {volume} {9}},\
  \bibinfo {pages} {158} (\bibinfo {year} {2023})}\BibitemShut {NoStop}%
\bibitem [{\citenamefont {Kim}\ and\ \citenamefont {Son}(2024)}]{DFTpseudoml2}%
  \BibitemOpen
  \bibfield  {author} {\bibinfo {author} {\bibfnamefont {R.}~\bibnamefont
  {Kim}}\ and\ \bibinfo {author} {\bibfnamefont {Y.-W.}\ \bibnamefont {Son}},\
  }\href {\doibase 10.1103/PhysRevB.109.045153} {\bibfield  {journal} {\bibinfo
   {journal} {Phys. Rev. B}\ }\textbf {\bibinfo {volume} {109}},\ \bibinfo
  {pages} {045153} (\bibinfo {year} {2024})}\BibitemShut {NoStop}%
\bibitem [{\citenamefont {Stippell}\ \emph {et~al.}(2024)\citenamefont
  {Stippell}, \citenamefont {Alzate-Vargas}, \citenamefont {Subedi},
  \citenamefont {Tutchton}, \citenamefont {Cooper}, \citenamefont {Tretiak},
  \citenamefont {Gibson},\ and\ \citenamefont {Messerly}}]{DFTpseudoml3}%
  \BibitemOpen
  \bibfield  {author} {\bibinfo {author} {\bibfnamefont {E.}~\bibnamefont
  {Stippell}}, \bibinfo {author} {\bibfnamefont {L.}~\bibnamefont
  {Alzate-Vargas}}, \bibinfo {author} {\bibfnamefont {K.~N.}\ \bibnamefont
  {Subedi}}, \bibinfo {author} {\bibfnamefont {R.~M.}\ \bibnamefont
  {Tutchton}}, \bibinfo {author} {\bibfnamefont {M.~W.}\ \bibnamefont
  {Cooper}}, \bibinfo {author} {\bibfnamefont {S.}~\bibnamefont {Tretiak}},
  \bibinfo {author} {\bibfnamefont {T.}~\bibnamefont {Gibson}}, \ and\ \bibinfo
  {author} {\bibfnamefont {R.~A.}\ \bibnamefont {Messerly}},\ }\href {\doibase
  https://doi.org/10.1016/j.aichem.2023.100042} {\bibfield  {journal} {\bibinfo
   {journal} {Artificial Intelligence Chemistry}\ }\textbf {\bibinfo {volume}
  {2}},\ \bibinfo {pages} {100042} (\bibinfo {year} {2024})}\BibitemShut
  {NoStop}%
\bibitem [{\citenamefont {Ghosh}\ and\ \citenamefont
  {Ghosh}(2023)}]{Ghosh2023}%
  \BibitemOpen
  \bibfield  {author} {\bibinfo {author} {\bibfnamefont {S.~K.}\ \bibnamefont
  {Ghosh}}\ and\ \bibinfo {author} {\bibfnamefont {D.}~\bibnamefont {Ghosh}},\
  }\href {\doibase 10.1063/5.0133399} {\bibfield  {journal} {\bibinfo
  {journal} {The Journal of Chemical Physics}\ }\textbf {\bibinfo {volume}
  {158}},\ \bibinfo {pages} {064108} (\bibinfo {year} {2023})},\ \Eprint
  {http://arxiv.org/abs/https://pubs.aip.org/aip/jcp/article-pdf/doi/10.1063/5.0133399/16734814/064108\_1\_online.pdf}
  {https://pubs.aip.org/aip/jcp/article-pdf/doi/10.1063/5.0133399/16734814/064108\_1\_online.pdf}
  \BibitemShut {NoStop}%
\bibitem [{\citenamefont {Arsenault}\ \emph {et~al.}(2014)\citenamefont
  {Arsenault}, \citenamefont {Lopez-Bezanilla}, \citenamefont {von
  Lilienfeld},\ and\ \citenamefont {Millis}}]{Arsenault2014}%
  \BibitemOpen
  \bibfield  {author} {\bibinfo {author} {\bibfnamefont {L.-F. m.~c.}\
  \bibnamefont {Arsenault}}, \bibinfo {author} {\bibfnamefont {A.}~\bibnamefont
  {Lopez-Bezanilla}}, \bibinfo {author} {\bibfnamefont {O.~A.}\ \bibnamefont
  {von Lilienfeld}}, \ and\ \bibinfo {author} {\bibfnamefont {A.~J.}\
  \bibnamefont {Millis}},\ }\href {\doibase 10.1103/PhysRevB.90.155136}
  {\bibfield  {journal} {\bibinfo  {journal} {Phys. Rev. B}\ }\textbf {\bibinfo
  {volume} {90}},\ \bibinfo {pages} {155136} (\bibinfo {year}
  {2014})}\BibitemShut {NoStop}%
\bibitem [{\citenamefont {Rigo}\ and\ \citenamefont
  {Mitchell}(2020)}]{Rigo2020}%
  \BibitemOpen
  \bibfield  {author} {\bibinfo {author} {\bibfnamefont {J.~B.}\ \bibnamefont
  {Rigo}}\ and\ \bibinfo {author} {\bibfnamefont {A.~K.}\ \bibnamefont
  {Mitchell}},\ }\href {\doibase 10.1103/PhysRevB.101.241105} {\bibfield
  {journal} {\bibinfo  {journal} {Phys. Rev. B}\ }\textbf {\bibinfo {volume}
  {101}},\ \bibinfo {pages} {241105} (\bibinfo {year} {2020})}\BibitemShut
  {NoStop}%
\bibitem [{\citenamefont {Sheridan}\ \emph {et~al.}(2021)\citenamefont
  {Sheridan}, \citenamefont {Rhodes}, \citenamefont {Jamet}, \citenamefont
  {Rungger},\ and\ \citenamefont {Weber}}]{Sheridan2021}%
  \BibitemOpen
  \bibfield  {author} {\bibinfo {author} {\bibfnamefont {E.}~\bibnamefont
  {Sheridan}}, \bibinfo {author} {\bibfnamefont {C.}~\bibnamefont {Rhodes}},
  \bibinfo {author} {\bibfnamefont {F.}~\bibnamefont {Jamet}}, \bibinfo
  {author} {\bibfnamefont {I.}~\bibnamefont {Rungger}}, \ and\ \bibinfo
  {author} {\bibfnamefont {C.}~\bibnamefont {Weber}},\ }\href {\doibase
  10.1103/PhysRevB.104.205120} {\bibfield  {journal} {\bibinfo  {journal}
  {Phys. Rev. B}\ }\textbf {\bibinfo {volume} {104}},\ \bibinfo {pages}
  {205120} (\bibinfo {year} {2021})}\BibitemShut {NoStop}%
\bibitem [{\citenamefont {Sturm}\ \emph {et~al.}(2021)\citenamefont {Sturm},
  \citenamefont {Carbone}, \citenamefont {Lu}, \citenamefont {Weichselbaum},\
  and\ \citenamefont {Konik}}]{Sturm2021}%
  \BibitemOpen
  \bibfield  {author} {\bibinfo {author} {\bibfnamefont {E.~J.}\ \bibnamefont
  {Sturm}}, \bibinfo {author} {\bibfnamefont {M.~R.}\ \bibnamefont {Carbone}},
  \bibinfo {author} {\bibfnamefont {D.}~\bibnamefont {Lu}}, \bibinfo {author}
  {\bibfnamefont {A.}~\bibnamefont {Weichselbaum}}, \ and\ \bibinfo {author}
  {\bibfnamefont {R.~M.}\ \bibnamefont {Konik}},\ }\href {\doibase
  10.1103/PhysRevB.103.245118} {\bibfield  {journal} {\bibinfo  {journal}
  {Phys. Rev. B}\ }\textbf {\bibinfo {volume} {103}},\ \bibinfo {pages}
  {245118} (\bibinfo {year} {2021})}\BibitemShut {NoStop}%
\bibitem [{\citenamefont {Walker}\ \emph {et~al.}(2022)\citenamefont {Walker},
  \citenamefont {Kellar}, \citenamefont {Zhang}, \citenamefont {Tam},\ and\
  \citenamefont {Moreno}}]{Walker2022}%
  \BibitemOpen
  \bibfield  {author} {\bibinfo {author} {\bibfnamefont {N.}~\bibnamefont
  {Walker}}, \bibinfo {author} {\bibfnamefont {S.}~\bibnamefont {Kellar}},
  \bibinfo {author} {\bibfnamefont {Y.}~\bibnamefont {Zhang}}, \bibinfo
  {author} {\bibfnamefont {K.-M.}\ \bibnamefont {Tam}}, \ and\ \bibinfo
  {author} {\bibfnamefont {J.}~\bibnamefont {Moreno}},\ }\href {\doibase
  10.3390/cryst12091269} {\bibfield  {journal} {\bibinfo  {journal} {Crystals}\
  }\textbf {\bibinfo {volume} {12}} (\bibinfo {year} {2022}),\
  10.3390/cryst12091269}\BibitemShut {NoStop}%
\bibitem [{\citenamefont {Murphy}(2022)}]{pml1Book}%
  \BibitemOpen
  \bibfield  {author} {\bibinfo {author} {\bibfnamefont {K.~P.}\ \bibnamefont
  {Murphy}},\ }\href {probml.ai} {\emph {\bibinfo {title} {Probabilistic
  Machine Learning: An introduction}}}\ (\bibinfo  {publisher} {MIT Press},\
  \bibinfo {year} {2022})\BibitemShut {NoStop}%
\bibitem [{\citenamefont {Goodfellow}\ \emph {et~al.}(2016)\citenamefont
  {Goodfellow}, \citenamefont {Bengio},\ and\ \citenamefont
  {Courville}}]{Goodfellow2016}%
  \BibitemOpen
  \bibfield  {author} {\bibinfo {author} {\bibfnamefont {I.}~\bibnamefont
  {Goodfellow}}, \bibinfo {author} {\bibfnamefont {Y.}~\bibnamefont {Bengio}},
  \ and\ \bibinfo {author} {\bibfnamefont {A.}~\bibnamefont {Courville}},\
  }\href@noop {} {\emph {\bibinfo {title} {Deep Learning}}}\ (\bibinfo
  {publisher} {MIT Press},\ \bibinfo {year} {2016})\ \bibinfo {note}
  {\url{http://www.deeplearningbook.org}}\BibitemShut {NoStop}%
\bibitem [{\citenamefont {Abadi}\ \emph {et~al.}(2015)\citenamefont {Abadi},
  \citenamefont {Agarwal}, \citenamefont {Barham}, \citenamefont {Brevdo},
  \citenamefont {Chen}, \citenamefont {Citro}, \citenamefont {Corrado},
  \citenamefont {Davis}, \citenamefont {Dean}, \citenamefont {Devin},
  \citenamefont {Ghemawat}, \citenamefont {Goodfellow}, \citenamefont {Harp},
  \citenamefont {Irving}, \citenamefont {Isard}, \citenamefont {Jia},
  \citenamefont {Jozefowicz}, \citenamefont {Kaiser}, \citenamefont {Kudlur},
  \citenamefont {Levenberg}, \citenamefont {Man\'{e}}, \citenamefont {Monga},
  \citenamefont {Moore}, \citenamefont {Murray}, \citenamefont {Olah},
  \citenamefont {Schuster}, \citenamefont {Shlens}, \citenamefont {Steiner},
  \citenamefont {Sutskever}, \citenamefont {Talwar}, \citenamefont {Tucker},
  \citenamefont {Vanhoucke}, \citenamefont {Vasudevan}, \citenamefont
  {Vi\'{e}gas}, \citenamefont {Vinyals}, \citenamefont {Warden}, \citenamefont
  {Wattenberg}, \citenamefont {Wicke}, \citenamefont {Yu},\ and\ \citenamefont
  {Zheng}}]{TensorFlow2015}%
  \BibitemOpen
  \bibfield  {author} {\bibinfo {author} {\bibfnamefont {M.}~\bibnamefont
  {Abadi}}, \bibinfo {author} {\bibfnamefont {A.}~\bibnamefont {Agarwal}},
  \bibinfo {author} {\bibfnamefont {P.}~\bibnamefont {Barham}}, \bibinfo
  {author} {\bibfnamefont {E.}~\bibnamefont {Brevdo}}, \bibinfo {author}
  {\bibfnamefont {Z.}~\bibnamefont {Chen}}, \bibinfo {author} {\bibfnamefont
  {C.}~\bibnamefont {Citro}}, \bibinfo {author} {\bibfnamefont {G.~S.}\
  \bibnamefont {Corrado}}, \bibinfo {author} {\bibfnamefont {A.}~\bibnamefont
  {Davis}}, \bibinfo {author} {\bibfnamefont {J.}~\bibnamefont {Dean}},
  \bibinfo {author} {\bibfnamefont {M.}~\bibnamefont {Devin}}, \bibinfo
  {author} {\bibfnamefont {S.}~\bibnamefont {Ghemawat}}, \bibinfo {author}
  {\bibfnamefont {I.}~\bibnamefont {Goodfellow}}, \bibinfo {author}
  {\bibfnamefont {A.}~\bibnamefont {Harp}}, \bibinfo {author} {\bibfnamefont
  {G.}~\bibnamefont {Irving}}, \bibinfo {author} {\bibfnamefont
  {M.}~\bibnamefont {Isard}}, \bibinfo {author} {\bibfnamefont
  {Y.}~\bibnamefont {Jia}}, \bibinfo {author} {\bibfnamefont {R.}~\bibnamefont
  {Jozefowicz}}, \bibinfo {author} {\bibfnamefont {L.}~\bibnamefont {Kaiser}},
  \bibinfo {author} {\bibfnamefont {M.}~\bibnamefont {Kudlur}}, \bibinfo
  {author} {\bibfnamefont {J.}~\bibnamefont {Levenberg}}, \bibinfo {author}
  {\bibfnamefont {D.}~\bibnamefont {Man\'{e}}}, \bibinfo {author}
  {\bibfnamefont {R.}~\bibnamefont {Monga}}, \bibinfo {author} {\bibfnamefont
  {S.}~\bibnamefont {Moore}}, \bibinfo {author} {\bibfnamefont
  {D.}~\bibnamefont {Murray}}, \bibinfo {author} {\bibfnamefont
  {C.}~\bibnamefont {Olah}}, \bibinfo {author} {\bibfnamefont {M.}~\bibnamefont
  {Schuster}}, \bibinfo {author} {\bibfnamefont {J.}~\bibnamefont {Shlens}},
  \bibinfo {author} {\bibfnamefont {B.}~\bibnamefont {Steiner}}, \bibinfo
  {author} {\bibfnamefont {I.}~\bibnamefont {Sutskever}}, \bibinfo {author}
  {\bibfnamefont {K.}~\bibnamefont {Talwar}}, \bibinfo {author} {\bibfnamefont
  {P.}~\bibnamefont {Tucker}}, \bibinfo {author} {\bibfnamefont
  {V.}~\bibnamefont {Vanhoucke}}, \bibinfo {author} {\bibfnamefont
  {V.}~\bibnamefont {Vasudevan}}, \bibinfo {author} {\bibfnamefont
  {F.}~\bibnamefont {Vi\'{e}gas}}, \bibinfo {author} {\bibfnamefont
  {O.}~\bibnamefont {Vinyals}}, \bibinfo {author} {\bibfnamefont
  {P.}~\bibnamefont {Warden}}, \bibinfo {author} {\bibfnamefont
  {M.}~\bibnamefont {Wattenberg}}, \bibinfo {author} {\bibfnamefont
  {M.}~\bibnamefont {Wicke}}, \bibinfo {author} {\bibfnamefont
  {Y.}~\bibnamefont {Yu}}, \ and\ \bibinfo {author} {\bibfnamefont
  {X.}~\bibnamefont {Zheng}},\ }\href {https://www.tensorflow.org/} {\enquote
  {\bibinfo {title} {{TensorFlow}: Large-scale machine learning on
  heterogeneous systems},}\ } (\bibinfo {year} {2015}),\ \bibinfo {note}
  {software available from tensorflow.org}\BibitemShut {NoStop}%
\bibitem [{\citenamefont {Haverkort}\ \emph {et~al.}(2012)\citenamefont
  {Haverkort}, \citenamefont {Zwierzycki},\ and\ \citenamefont
  {Andersen}}]{Haverkort_2012}%
  \BibitemOpen
  \bibfield  {author} {\bibinfo {author} {\bibfnamefont {M.~W.}\ \bibnamefont
  {Haverkort}}, \bibinfo {author} {\bibfnamefont {M.}~\bibnamefont
  {Zwierzycki}}, \ and\ \bibinfo {author} {\bibfnamefont {O.~K.}\ \bibnamefont
  {Andersen}},\ }\href {\doibase 10.1103/PhysRevB.85.165113} {\bibfield
  {journal} {\bibinfo  {journal} {Phys. Rev. B}\ }\textbf {\bibinfo {volume}
  {85}},\ \bibinfo {pages} {165113} (\bibinfo {year} {2012})}\BibitemShut
  {NoStop}%
\bibitem [{\citenamefont {Geron}(2019)}]{HandsOnML}%
  \BibitemOpen
  \bibfield  {author} {\bibinfo {author} {\bibfnamefont {A.}~\bibnamefont
  {Geron}},\ }\href@noop {} {\emph {\bibinfo {title} {Hands-On Machine Learning
  with Scikit-Learn, Keras, and TensorFlow: Concepts, Tools, and Techniques to
  Build Intelligent Systems}}},\ \bibinfo {edition} {2nd}\ ed.\ (\bibinfo
  {publisher} {O'Reilly Media, Inc.},\ \bibinfo {year} {2019})\BibitemShut
  {NoStop}%
\bibitem [{\citenamefont {Karrasch}\ \emph {et~al.}(2008)\citenamefont
  {Karrasch}, \citenamefont {Hedden}, \citenamefont {Peters}, \citenamefont
  {Pruschke}, \citenamefont {Schönhammer},\ and\ \citenamefont
  {Meden}}]{Karrasch2008}%
  \BibitemOpen
  \bibfield  {author} {\bibinfo {author} {\bibfnamefont {C.}~\bibnamefont
  {Karrasch}}, \bibinfo {author} {\bibfnamefont {R.}~\bibnamefont {Hedden}},
  \bibinfo {author} {\bibfnamefont {R.}~\bibnamefont {Peters}}, \bibinfo
  {author} {\bibfnamefont {T.}~\bibnamefont {Pruschke}}, \bibinfo {author}
  {\bibfnamefont {K.}~\bibnamefont {Schönhammer}}, \ and\ \bibinfo {author}
  {\bibfnamefont {V.}~\bibnamefont {Meden}},\ }\href {\doibase
  10.1088/0953-8984/20/34/345205} {\bibfield  {journal} {\bibinfo  {journal}
  {Journal of Physics: Condensed Matter}\ }\textbf {\bibinfo {volume} {20}},\
  \bibinfo {pages} {345205} (\bibinfo {year} {2008})}\BibitemShut {NoStop}%
\end{thebibliography}%
\end{document}